\documentclass[journal,onecolumn]{IEEEtran}
\ifCLASSINFOpdf
\else
\fi

\usepackage{graphicx}
\usepackage[graphicx]{realboxes}
\usepackage{cite}
\usepackage{multirow}
\usepackage{enumitem}
\newlist{tabitemize}{itemize}{1}
\setlist[tabitemize]{label=\textbullet,nosep,after=\strut,align=parleft,leftmargin=*,}

\usepackage[T1]{fontenc}
\usepackage{bm}
\usepackage{vcell}

\def\*#1{\mathbf{#1}}

\begin{document}
%
\title{Synthetic CT image generation from CBCT:\\A Systematic Review}
\author{Alzahra~Altalib,~Scott~McGregor,~Chunhui~Li,~Alessandro~Perelli 
\thanks{This work did not involve human subjects or animals in its research. A. Altalib, C. Li, and A. Perelli are with the School of Science and Engineering, Center of Medical Engineering and Technology, University of Dundee, DD1 4HN, UK, email: \{2600129, c.li, aperelli001\}@dundee.ac.uk. A. Perelli is supported by the Royal Academy of Engineering under the RAEng / Leverhulme Trust Research Fellowships program (award LTRF-2324-20-160). S. McGregor is with the Library \& Learning Centre, University of Dundee, DD1 4HN, UK, e-mail: S.V.Mcgregor@dundee.ac.uk.} 
\thanks{Manuscript accepted 15 January, 2025.}}

\markboth{}%
{A. Altalib \MakeLowercase{\textit{et al.}}: Synthetic high-quality CT image generation from low-quality Cone-Beam CT}

\maketitle

\begin{abstract}
The generation of synthetic CT (sCT) images from cone-beam CT (CBCT) data using deep learning methodologies represents a significant advancement in radiation oncology. This systematic review, following PRISMA guidelines and using the PICO model, comprehensively evaluates the literature from 2014 to 2024 on the generation of sCT images for radiation therapy planning in oncology. A total of 35 relevant studies were identified and analyzed, revealing the prevalence of deep learning approaches in the generation of sCT. This review comprehensively covers synthetic CT generation based on CBCT and proton-based studies. Some of the commonly employed architectures explored are convolutional neural networks (CNNs), generative adversarial networks (GANs), transformers, and diffusion models. Evaluation metrics including mean absolute error (MAE), root mean square error (RMSE), peak signal-to-noise ratio (PSNR) and structural similarity index (SSIM) consistently demonstrate the comparability of sCT images with gold-standard planning CTs (pCT), indicating their potential to improve treatment precision and patient outcomes. Challenges such as field-of-view (FOV) disparities and integration into clinical workflows are discussed, along with recommendations for future research and standardization efforts. In general, the findings underscore the promising role of sCT-based approaches in personalized treatment planning and adaptive radiation therapy, with potential implications for improved oncology treatment delivery and patient care.
\end{abstract}


\IEEEpeerreviewmaketitle

\section{Introduction}

\IEEEPARstart{C}{one} Beam Computed Tomography (CBCT) is a specific type of CT where a cone-shaped X-ray beam is used for the generation of 3D images \cite{b25}. Such images are high-resolution images generated using the unique anatomy of a patient. CBCT differs from conventional CT scanners, which often employ fan-shaped X-ray beams. CBCT functions by rotating the source and detector in a single trial on the patient’s body surface, allowing the generation of a volumetric dataset in a single scan. A CBCT system comprises several components, including an X-ray source, a flat panel detector, and a gantry for the rotation of the X-ray tube. During the scanning process, a cone-shaped beam penetrates the patient’s body,causing attenuation and dispersion of the beams. These attenuated and dispersed beams are transformed into images using reconstruction techniques, resulting in 2D and 3D images \cite{b12, b18}.

CBCT systems offer numerous advantages in radiotherapy compared to conventional systems. The high-resolution 3D image construction capability allows us the capture of finer anatomical details. Additionally, by collecting volumetric data in a single rotation, CBCT reduces acquisition time and X-ray exposure. This lower radiation dose compared to conventional CT makes it suitable for more frequent imaging and diagnosis \cite{b30}. Such benefits have enabled the application of CBCT in various fields, including diagnostic imaging, treatment planning, and intraoperative guidance. CBCT can precisely localize tumors and surrounding organs, enabling frequent target localization during treatment delivery. This capability allows medical personnel to capture 3D volumetric images before each treatment session, adjust the target volume position, and ensure accurate radiation delivery. 

Image-guided radiotherapy (IGRT) involves the use of CBCT which provides information about an individual's anatomy prior to therapy. IGRT is the observation of the patient's morphological differences between the time of the initial CT scan and subsequent therapy sessions (fractionally distributed), and it may be possible to alter therapy based on the detected variations \cite{b8}. One restriction to the practical application of CBCT in extracting patient information is because the visual appearance of CBCT scans is considered much worse than planned CT examinations with respect to contrast-to-noise ratios and the presence of imaging artifacts including streaking, which can be up to 50\% more prevalent in CBCT scans \cite{b7}. X-ray scatter or internal motion are among the causes of streak abnormalities in abdominal imaging. This has sparked an increasing interest in developing strategies to improve the quality of CBCT scans to match that of CT \cite{b40}. 

The most well-established approach for producing synthetic CT (sCT) images with CT images is deformable image registration (DIR), which involves deforming the planned CT to fit the geometry of the CBCT \cite{b9}. The fundamental problem of DIR-based techniques is that the non-deformable alteration is not taken into account during the scans. For example, instances of lung compression and variation in air volumes and placements of the gastrointestinal (GI) are not taken into account \cite{b29}. Post-processing techniques can mitigate gross anatomical mismatches to some extent. But most DIR methods are too slow for real-time applications, which require computational performance equivalent to that of graphics processing units (GPUs) \cite{b6}. 
An alternative approach involves adjusting the scattering levels in the CBCT images.  

Over the past two decades, deep learning (DL) has emerged as a promising solution for healthcare image synthesis applications, including CBCT-to-CT conversion. Compared to traditional methods, data-driven DL approaches offer superior performance and adaptability to unexplored datasets. However, the need for large-scale data collection and curation remains a significant challenge \cite{b11}.

DL has revolutionized medical imaging by learning complex representations from raw data, bypassing the reliance on hand-crafted features. This capability makes DL ideal for enhancing CBCT images in applications such as Adaptive Radiation Therapy (ART), where accurate imaging is critical to modifying treatment plans in real time. Traditional image processing methods struggle with the complexity of medical images, underscoring the need for advanced DL techniques that can automatically improve image quality and enhance the results of ART \cite{b49}. An alternative approach involves making adjustments to the levels of scattering in the CBCT images, which further contributes to improving image quality and reducing artifacts.

In recent years, extensive research has focused on DL for image transformation, including hundreds of studies on CBCT-to-CT conversion. Several DL frameworks have been developed for this purpose, including U-Nets, paired pix2pix Generative Adversarial Networks (GANs) \cite{b45, b44} and paired and unpaired cycle-consistent GANs (CycleGANs) \cite{b46, b47}. Data-driven approaches such as these enable the translation of images between two imaging modalities. Unpaired CycleGANs are particularly advantageous in CBCT-to-CT synthesis because they do not require structurally connected paired data. This approach is especially relevant because simultaneous acquisitions of CBCT and CT scans are often impractical, and closely spaced scans reduce structural variations \cite{b28}.

In the following section, we review the main DL architectures commonly applied to CBCT to CT image synthesis. 

\subsection{Deep Learning models for CBCT-to-CT Image Synthesis}

\subsubsection{\textbf{Convolutional Neural Networks (CNNs)}} they have been extensively applied in medical image analysis, particularly in tasks that require high precision, such as noise reduction and artifact removal. These techniques are particularly critical in ART, where accurate patient anatomy representation is necessary for real-time treatment adjustments \cite{b50}. Such an architecture allows the extraction of intricate details from within the medical images. Such details can be in the form of edges, gradients, and textures that can help refine the quality of CBCT images \cite{b49}. The capability of CNNs to capture both local and global image features makes them particularly useful in reconstructing CBCT scans, which are often plagued by noise and scatter \cite{b50}.

\begin{figure}[!h]
\centering
\includegraphics[width=.75\linewidth]{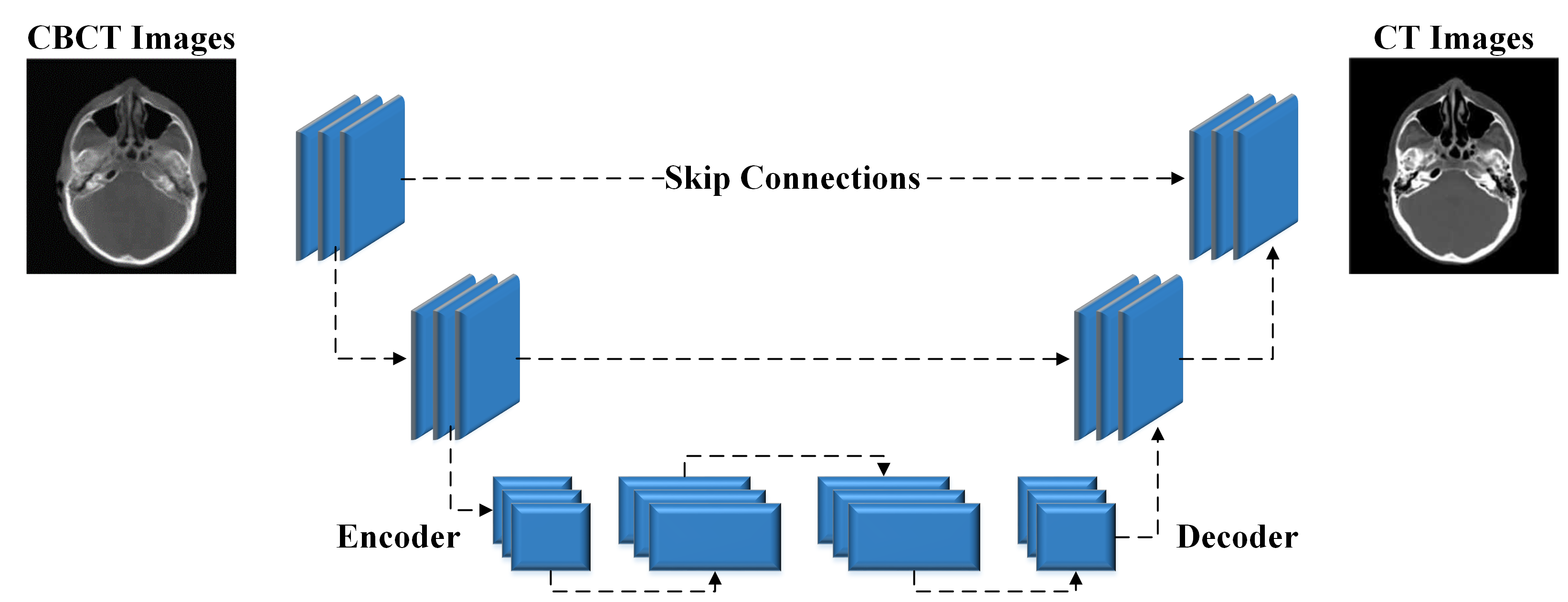}
\caption{Schematic diagram of U-Net for CBCT-to-CT image translation. The U-Net architecture features an encoder-decoder design with skip connections to preserve spatial details. The input CBCT images are compressed by the encoder and then reconstructed back to corresponding CT images. (CT and CBCT images taken from \cite{b48}.)}
\label{U-Net_Architectures}
\end{figure}

\textbf{U-Net architecture} is widely regarded as one of the most effective CNN models for the segmentation and enhancement of medical images. Its encoder-decoder structure allows the model to compress the image information into a latent space before reconstructing it with finer details via skip connections between the corresponding encoder and decoder layers \cite{b49}. U-Net has demonstrated significant success in tasks such as CBCT image correction due to its ability to produce pixel-level predictions, which are critical for applications in ART. The use of skip connections ensures that spatial and contextual information is preserved within the network, allowing more accurate reconstructions \cite{b50}. The input of the U-Net is a batch of CBCT images and the output CT images; during the training most often the Mean Squared error (MSE) metrics between CBCT and CT pairs images is minimized. The standard U-Net architecture is shown in Fig. \ref{U-Net_Architectures}.

\subsubsection{\textbf{Generative Adversarial Networks (GANs)}} 
\begin{figure}[!h]
\centering
\includegraphics[width=.6\linewidth]{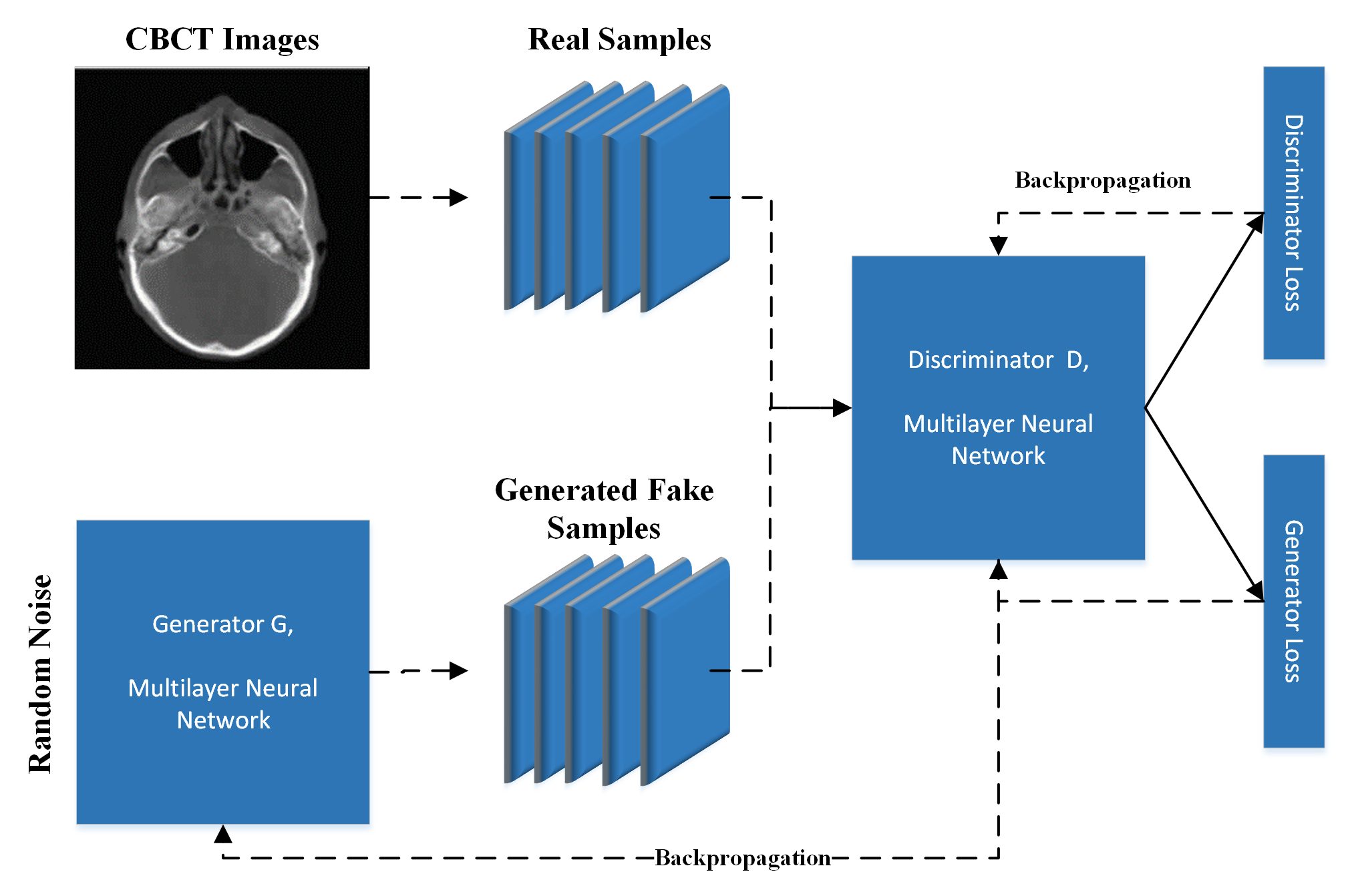}
\caption{Generative Adversarial Network (GAN) framework of CBCT image synthesis. The architecture includes a generator (G), comprising as multilayer neural network, that generates CBCT images out of random noise. The discriminator (D) is a multilayer neural net, the input will be images, and its targets are binary answers, real or fake. (CBCT image taken from \cite{b48}.)}
\label{GAN_Architectures}
\end{figure}

this family of network introduce a dual-network framework that involves a generator and a discriminator. The generator attempts to create synthetic images that resemble real ones, while the discriminator evaluates the quality of these images, classifying them as real or synthetic \cite{b49}. GANs were first introduced in \cite{b51} and have since been widely adopted in medical image synthesis due to their ability to improve the realism of generated images. This adversarial process significantly improves the quality of CBCT images, especially in cases where paired datasets are unavailable. GANs have been successfully applied in CBCT image enhancement, where their ability to handle unpaired data sets makes them particularly useful \cite{b51}. The adversarial loss used in GANs encourages the generation of high-fidelity images, which is essential for accurate dose calculations in ART. The standard GAN architecture is shown in Fig. \ref{GAN_Architectures}.

\begin{figure}[!h]
\centering
\includegraphics[width=.75\linewidth]{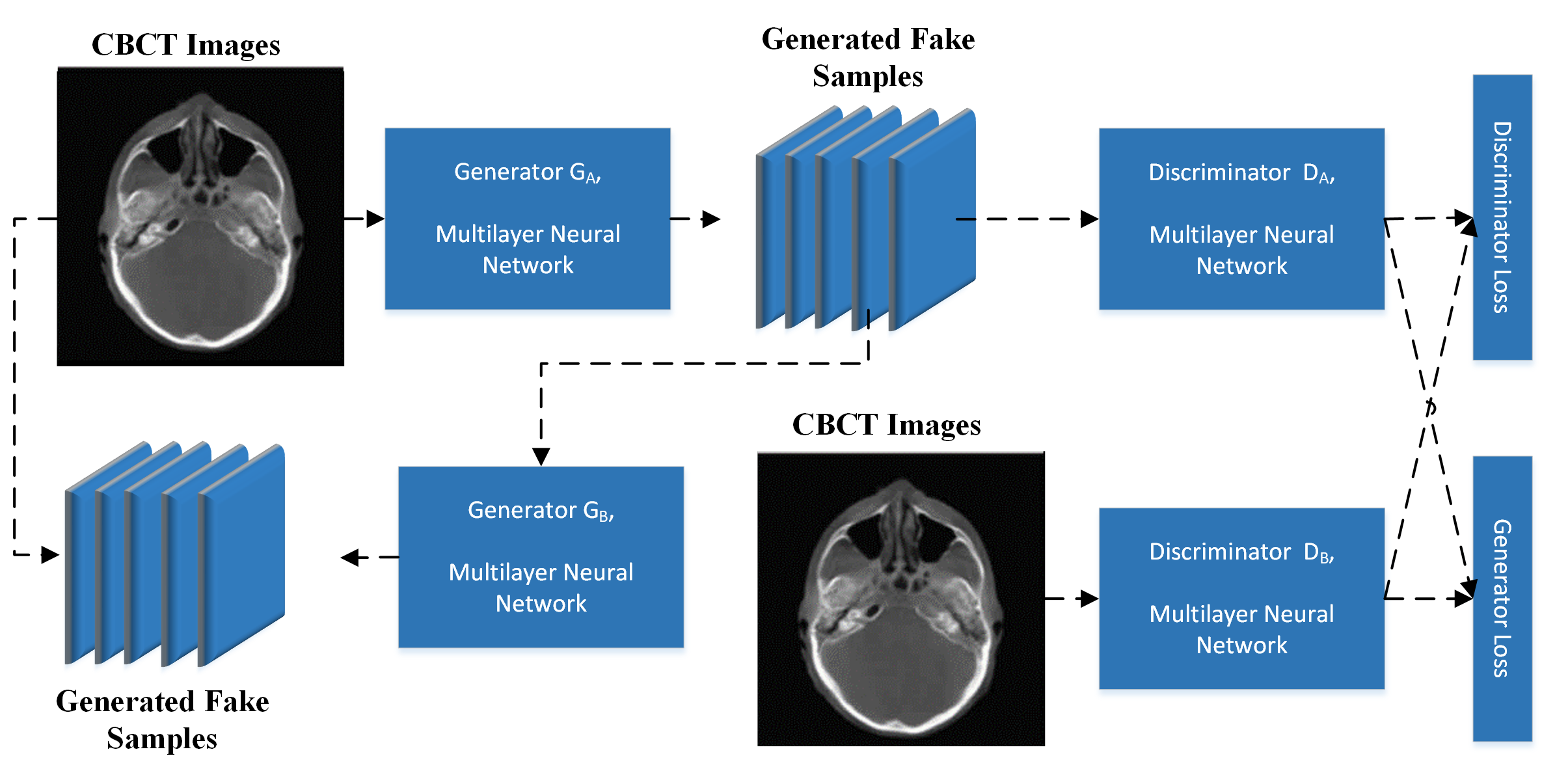}
\caption{Cyclic GAN architecture for CBCT image synthesis. The architecture includes two generator modules ($G_A$ and $G_B$) to transfer between domains A and B, as well as discriminator modules $D_A$ and $D_B$. The Generators are trained to produce synthetic CBCT images from real ones. The discriminators judge the realism of generated images with a cycle consistency enforced to optimize the model. (CT and CBCT images taken from \cite{b48}.)}
\label{Cyclic_GANs_Architectures}
\end{figure}

\textbf{Cycle-GAN architecture} extends the GAN framework by introducing a loss of cycle consistency, ensuring that transformations between domains (e.g. from CBCT to CT and back) preserve anatomical structures \cite{b49}. This architecture is effective for CBCT enhancement when paired data is unavailable. Cycle-GANs are useful in ART, where anatomical precision is crucial for real-time dose recalculations. Despite their computational complexity, Cycle-GANs have shown promising results in improving the usability of CBCT images for clinical applications \cite{b51}. However, they often show instabilities during training due to adversarial, min-max optimization. The CyclicGAN architecture is shown in Fig. \ref{Cyclic_GANs_Architectures}.

\subsubsection {\textbf{Transformer-based network (TransCBCT)}} 
\begin{figure}[!h]
\centering
\includegraphics[width=.75\linewidth]{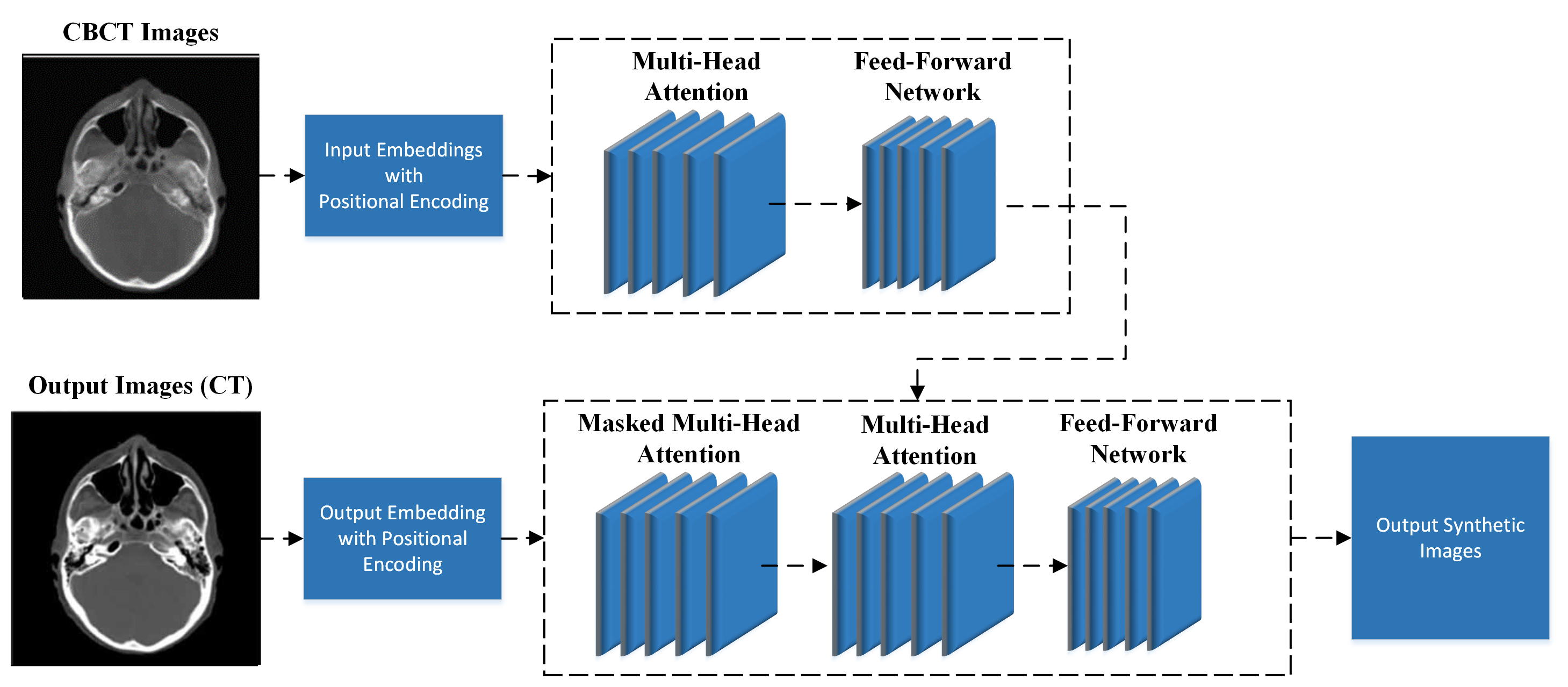}
\caption{Transformer-Based Architecture for CBCT to CT image synthesis (TransCBCT). The network leverages self-attention mechanisms, including multi-head and masked multi-head attention, to capture long-range dependencies in CBCT images. Positional encoding is applied to input and output embeddings, allowing the model to synthesize CT images by processing and transforming CBCT inputs through a sequence of attention layers and feed-forward networks. (CT and CBCT images taken from \cite{b48}.)}\label{TransCBCT_Architectures}
\end{figure}

it uses a self-attention mechanism to capture local and global dependencies contained in the images. Compared to CNN-based architecture, the TransCBCT model can learn long-range relationships.This makes it highly suitable for tasks such as CT image synthesis. The models can learn from the spatial arrangement of anatomical structures contained in the brain images. TransCBCT consists of an encoder-decoder layer, self-attention of multiple heads, and positional encoding to effectively model the spatial information from the CBCT scans \cite{b3}. The architecture of the model is depicted in Fig. \ref{TransCBCT_Architectures}.

\begin{table*}[!ht]
\centering
\caption{Strengths and weaknesses of the deep learning architectures used for synthetic CT image generation.}\label{tab:deep_learning_models}
\begin{tabular} {p{65pt}p{55pt}p{180pt}p{140pt}}
\hline
\vspace{.02cm}\textbf{Main Architecture} & \textbf{Specific Network} & \vspace{.02cm}\textbf{Benefits and Strengths} & \vspace{.02cm}\textbf{Limitations}  \\
\hline
\vspace{.6cm}Convolutional \newline Networks &
\vspace{.6cm}U-Net
&
\begin{tabitemize}
\item Effective for image segmentation.
\item Handles small datasets.
\item Symmetric design aids localization tasks.
\item Simplest implementation.
\item Stable convergence.
\item Fastest training.
\end{tabitemize} & 
\begin{tabitemize}
\item Paired data only.
\item Anatomic misalignments reduce model accuracy.
\end{tabitemize} \\
\hline
\vspace{1.6cm}Generative Networks& 
\vspace{.6cm}GAN
& 
\begin{tabitemize}
\item Highly successful in generating realistic images.
\item Can be used for image synthesis and style transfer.
\item Paired or unpaired training.
\item Improved image realism due to adversarial loss.
\item Model tunability.
\end{tabitemize} &
\begin{tabitemize}
\item Moderate implementation difficulty. 
\item Unstable convergence.  
\item Slower training. 
\end{tabitemize} \\
&
\vspace{.6cm}Cycle-GAN	
&   
\begin{tabitemize}
\item Performs unpaired image-to-image translation.
\item Useful when paired data is unavailable.
\item Paired or unpaired training.
\item Model tunability.
\item Good structure preservation.  
\end{tabitemize} &
\begin{tabitemize}
\item Complex implementation.
\item Unstable convergence.
\item Slowest training.
\item High hardware requirements.
\end{tabitemize}  \\
\hline
\vspace{.2cm}Transformer-Based Models &
\vspace{.3cm}TransCBCT  
& 
\begin{tabitemize}
\item Captures long-range dependencies. 
\item No bias toward local features. 
\item Efficient for processing large image datasets.	 
\end{tabitemize} & 
\begin{tabitemize}
\item Requires large datasets. 
\item High computational cost. 
\item Sensitive to image resolution.
\end{tabitemize} \\
\hline
\vspace{.3cm}Diffusion Models &
\vspace{.3cm}DDPM	
&
\begin{tabitemize}
\item Generates high-quality images. 
\item Robust to noise. 
\item Stable training.	
\end{tabitemize} &
\begin{tabitemize}
\item Computationally expensive. 
\item Slow generation process in testing.
\item High hardware requirements.
\end{tabitemize} \\
\hline
\end{tabular}
\end{table*}

\subsubsection{\textbf{Conditional Denoising Diffusion Probabilistic Model (DDPM)}} 
\begin{figure}[!h]
\centering
\includegraphics[width=.75\linewidth]{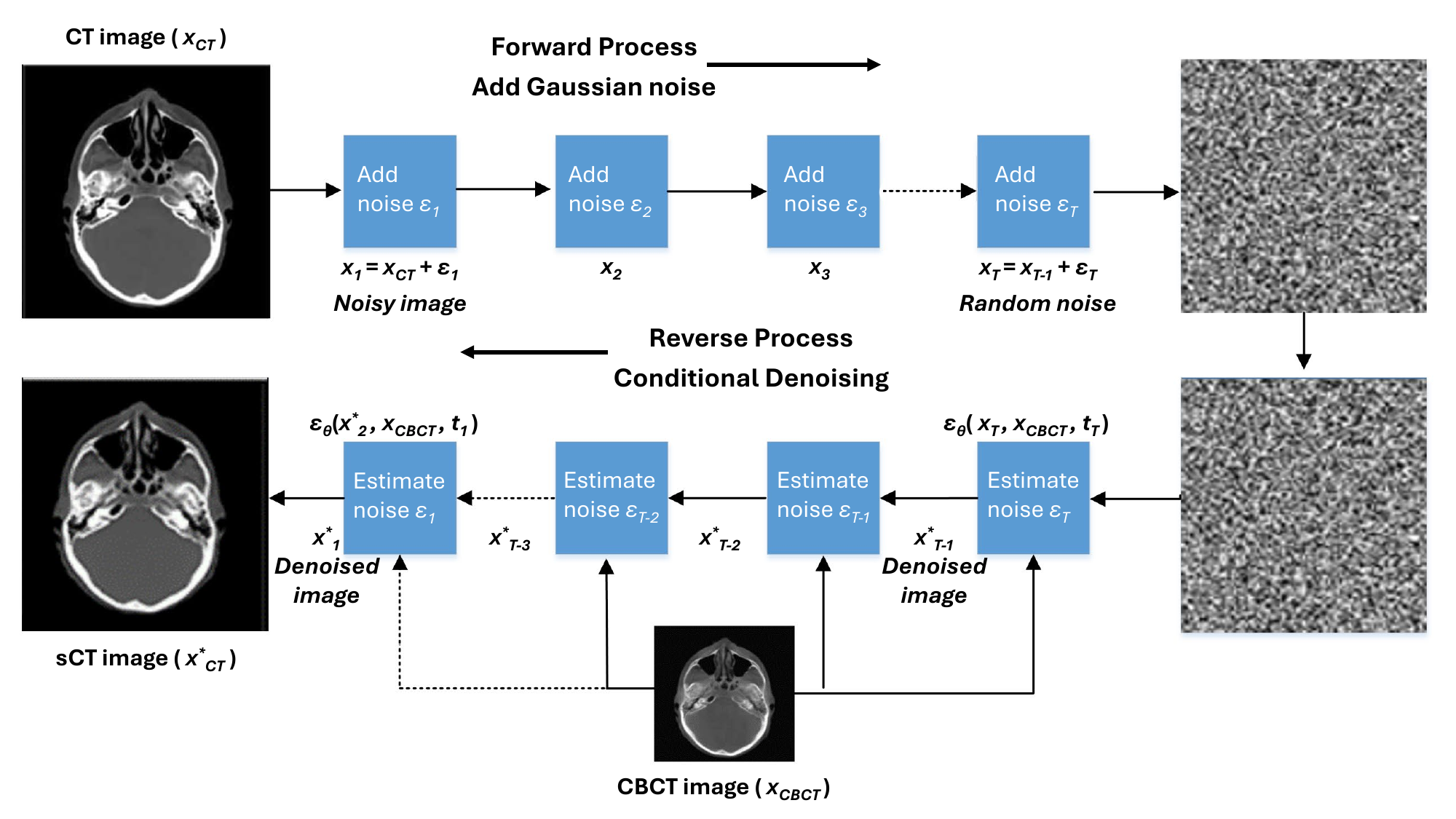}
\caption{Denoising Diffusion Probabilistic Model (DDPM) for CBCT-to-CT image synthesis. The model employs a diffusion-based framework where Gaussian noise is progressively added to the CT image $\*x_{CT}$ over $T$ time steps. This results in intermediate noisy representations denoted as $\*x_1, \*x_2, \ldots, \*x_T$. In the reverse process, the model iteratively estimates and removes noise using a neural network with parameters $\bm\mu$, conditioned on the CBCT image $\*x_{CBCT}$. (CT and CBCT images taken from \cite{b48}.)}
\label{DDPM_Architectures}
\end{figure}

is a type of generative model that aims to denoise the image using an iterative approach. The training consists of two steps: the forward process, in which a deterministic amount of Gaussian noise is progressively added to the input image until reaching completely random noise, and the reverse process, which aims to learn the noise at each step using a temporally dependent DL network. DDPM has significant applications in medical images, especially those related to the generation of CT images. This ability leads to the analysis of complex distributions and the generation of high-quality images \cite{b23}. To improve the consistency of the DDPM generative capability with actual data, conditioning the reverse process on real external data ensures that the model generates synthetic CT images based on specific input conditions and anatomical features derived from real acquired CBCT images. The model architecture is depicted in Fig. \ref{DDPM_Architectures}.

During the forward process, the CT image $\*x_{CT}$ is used as input where at each iteration step $t = 1,\ldots, T$, the Gaussian noise $\epsilon_t\sim\mathcal{N}(0, \sigma_t^2)$ is added iteratively, to generate noisy images 
\begin{equation}
    \*x_t = \*x_{t-1} + \bm\epsilon_t, \quad t = 1,\ldots, T
\end{equation}
until the noisy image $x_T$ is obtained that contains no discernible features of the original CT image and represents a random noise-like structure. During the reverse process, the conditional denoising is performed. The model starts by removing the noise added during the forward process to reconstruct a sCT image $\*x^*_{CT}$ that has been conditioned on the acquired CBCT image $\*x_{CBCT}$. In the reverse process, the noisy images $x_T$ and $x_{CBCT}$ are passed to the model. At each step, the noise $\bm\epsilon$ added during the forward process is estimated through a time dependent neural network with parameters $\bm\mu$, conditioned on the CBCT image as
\begin{equation}
    \bm\epsilon_{\bm\theta}(\*x^*_{t+1}, \*x_{CBCT}, t_t), \quad t = T, T-1, \ldots, 1
\end{equation}
and the denoised intermediate CT images $\*x^*_t, \; t = T, \ldots, 1$ are generated as a weighted linear combination 
\begin{equation}
    \*x^*_{t} = \*x^*_{t+1} - s_t\left(\bm\epsilon_{\bm\theta}(\*x^*_{t+1}, \*x_{CBCT}, t_t)\right)
\end{equation}
where $s_t$ is a scalar term dependent on $t$. The estimated noise is gradually removed, thus producing a less noisy version of the original CT image at each step and $\*x^*_{CT}$ as output.

Each of the four main architectures depicted above has its merits and limitations in terms of synthetic CT image generation. A summary of the strengths and weaknesses associated with each of these models is depicted in Table \ref{tab:deep_learning_models}.

\subsection{Model Optimization and Loss Functions} 

The optimization process in DL models revolves around minimizing the difference between the predicted and actual output. This is typically achieved by defining a loss function that quantifies the error in the predictions of the model. Common loss functions, such as mean absolute error (MAE) and mean squared error (MSE), are employed to measure pixel-level discrepancies between synthetic and ground-truth images \cite{b49}. Furthermore, gradient descent optimization adjusts the model parameters iteratively, reducing the loss and improving the image quality of the CBCT scans \cite{b49}. This optimization process is crucial for generating high-quality sCT images that can be used for dose recalculations in ART.

\subsection{Pre-processing Techniques for Medical Image Analysis}

Pre-processing is a critical step in medical image analysis, aimed at overcoming limitations associated with medical data collection and enhancing the performance of downstream tasks, such as segmentation, classification, and detection, in clinical applications. Below are some key pre-processing techniques commonly used in this field:

\textbf{Registration}: Image registration involves aligning images from different time points, modalities, or perspectives into a common coordinate system to facilitate accurate comparison and integration. This technique is vital for tasks such as longitudinal studies in which images taken from the same patient at different times must be precisely aligned. Registration methods can be rigid or non-rigid, with the latter accounting for more complex deformations like organ movement. Methods based on mutual information and, more recently, deep learning have shown significant improvements in both the accuracy and speed of registration \cite{b53 , b54}.

\textbf{Data Augmentation}: Data augmentation is a method to artificially increase the size of the training dataset by applying random transformations such as rotation, scaling, flipping, and adding noise. This technique is critical for deep learning applications, where large datasets are needed to prevent over-fitting and improve generalization. In medical imaging, where obtaining new labeled data can be difficult due to high costs and privacy concerns, data augmentation serves as an efficient solution to simulate variability and improve model performance \cite{b55}.

\textbf{Paired and Unpaired Data}: Misalignment Between Labels and Images: In supervised learning, paired data, where each image has a corresponding label, are necessary for training. However, in real-world medical datasets, obtaining perfectly paired data is challenging due to issues such as annotation errors or imaging artifacts, which can introduce noise into the training process. Techniques like weakly-supervised and unsupervised learning have emerged to address these challenges, allowing the use of unpaired data. Generative adversarial networks (GANs) have been especially useful in mitigating the misalignment between labels and images, providing effective solutions for training with unpaired data \cite{b56, b47}.

\textbf{2D and 3D Datasets}: Medical images can be represented in either 2D, such as X-rays or individual slices of CT or MRI scans, or in 3D, such as volumetric CT and MRI datasets. Pre-processing for 3D datasets typically involves volumetric transformations and requires significantly more computational power compared to 2D datasets. However, while 3D datasets offer richer spatial information, 2D data are easier to handle computationally. Hybrid approaches that combine 2D and 3D information, such as using 2D slices along with contextual 3D information, have been developed of both types of data \cite{b50}.

\subsection{Research Question and Aims}
\begin{figure}[!h]
\centering
\includegraphics[width=.75\linewidth]{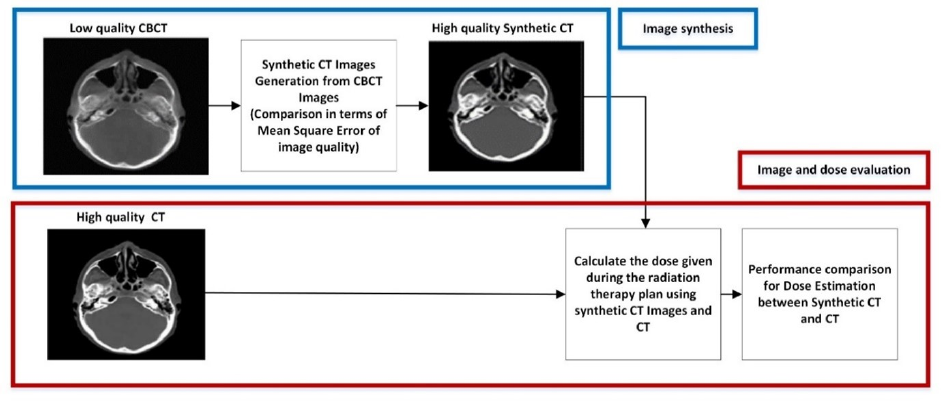}
\caption{Schematic diagram of the image synthesis approach  to generate high quality synthetic CT images from CBCT. (CT and CBCT images taken from \cite{b48}.)}
\label{fig_diag}
\end{figure}

The goal of this review is to determine whether the integration of deep learning methodologies in the creation of CBCT data-based sCT images can improve the precision and effectiveness of radiation therapy planning, while also potentially lowering exposure. There is no systematic review of articles related to the generation of CBCT sCT. The review focuses primarily on papers that follow the image synthesis approach based on the image synthesis flow provided in Fig. \ref{fig_diag}. Specifically, it focuses on the deep learning methodologies employed to generate high-quality CT images from low-quality CBCT images. The objective is to synthesize high-quality CT images from CBCT inputs using advanced deep learning techniques.

\subsubsection {Research Question}
This review focuses on answering this question: Can the integration of deep learning methodologies for the generation of sCT images from CBCT data enhance the accuracy and effectiveness of radiation therapy planning in oncology and potentially reduce exposure?

\subsubsection{Study Aims}
This systematic review aims to collect and analyze data on CBCT sCT generation using deep learning and neural network methods.

\section{Methodology}
In this section, the discussion begins with an exploration of the search strategy employed for the review. Subsequently, the methodological steps taken to conduct the review are described in detail.

\subsection{Search Strategy}
This systematic search followed the PRISMA statement \cite{b22} and used the PICO model. See Table \ref{Search_Strategy} to find relevant literature. PubMed, MEDLINE, and Web of Science (WOS) databases were searched between 2014 and 2024 following the defined study criteria. Data collection continued until March 17, 2024, to ensure the inclusion of all pertinent studies. The search strategy used a combination of phrases and keywords relevant to the research question to guarantee comprehensive coverage. These included'synthetic CT,' 'Computed Tomography,' 'cone beam computed tomography,' 'dose calculation,' and synonyms such as 'CBCT.' Boolean operators (AND, OR)" were utilized search for different databases appendix to efficiently filter and merge search keywords.

\begin{table}[h!]
\centering
\caption{The PICO framework for systematic reviews}
\label{Search_Strategy}
\begin{tabular}{p{40pt}p{180pt}}
\hline
Population 	&	All patients underwent definitive oncology planning.	\\
\hline
Intervention	&	DL OR Deep learning OR Convolutional Neural Network OR CNN, CBCT OR Cone beam Computed tomography, Imaging Reconstruction or IR, Unsupervised Deep Learning OR UDL, Dose Estimations OR DE, Medical Imaging OR MI, Artifact Reduction OR AR, Dose Estimation OR DE, radiotherapy	\\
\hline
Comparison	&	CT OR computed tomography	\\
\hline
Outcome 	&	Sensitivity, specificity, accuracy	\\
\hline
\end{tabular}
\end{table}

\subsection{Inclusion and exclusion criteria of the study}

The studies were included if they:
\begin{itemize}
    \item Emphasize the creation of sCT images using CBCT datasets.
    \item sCT images must be used in studies to calculate doses in the framework of definitive oncology planning.
    \item Full-text peer-reviewed journal articles are included.
    \item Articles published in the English language.
\end{itemize}

The studies were excluded if they:
\begin{itemize}
    \item Related to other modalities rather than CBCT for the generation of sCT.
    \item Studies not involving dose calculation strategies for definitive oncology planning. 
    \item Studies not published in the English language.
    \item Student dissertations, review articles, unpublished data and data that are not accessible to online databases.  
\end{itemize}

\subsection{Study Selection}
Following the systematic search, articles are screened according to their titles and abstracts to identify potentially relevant studies. After undergoing an initial screening process, the articles undergo a full text review, where their suitability for inclusion in the systematic review was thoroughly evaluated. Inclusion / exclusion criteria were strictly applied during the screening process, with reasons for exclusion documented at the full text stage for transparency and reproducibility. 

\subsection{Data Extraction}
After retrieving the results from the database search, the files were transferred to a reference manager to consolidate all search outcomes. The reference manager used in this systematic review was EndNote. At this stage, duplicate records were removed. Two reviewers, Alzahra Altalib (AA) and Alessandro Perelli (AP), independently assessed and applied the predetermined eligibility criteria to all studies selected for inclusion in the systematic review. In cases of disagreement, a discussion took place between the two to reach a consensus.
Data extraction has been carried out from selected articles based on a standardized data extraction approach. The extracted information includes publication details such as year, authors, and country of origin, as well as specific details related to CBCT acquisition parameters (e·g·, image resolution, field of view, scanning protocol), synthetic CT generation methodologies (e·g·, image registration, segmentation, machine learning-based approaches), dose calculation algorithms utilized (e·g·, analytical, Monte Carlo), and the validation techniques employed.

\subsection{Quality Assessment}
To evaluate the methodological quality of the included studies, two researchers, AA and PA, used the Quality Assessment of Diagnostic Accuracy Studies-2 (QUADAS-2) tool \cite{b42}. QUADAS-2 was utilized to assess the risk of bias. This tool provides structured frameworks for assessing the risk of bias, the methodological rigor, and the overall quality of each study included in the systematic review.

\subsection{Data Synthesis}
The synthesis data was examined and presented to identify general patterns, advantages, limitations, and deficiencies in research related to CBCT synthetic CT production for dose estimation in radiation treatment. The studies were classified according to the sCT generation methodology used, such as voxel-based techniques, atlas-based techniques, and deep learning-based methods. Any variability among included research, such as variations in research design, patient demographics, or validation methodologies, is acknowledged and taken into account when interpreting the results. Finally, the study's consequences for clinical treatment and future research prospects were reviewed.

\subsection{Results} 
\begin{figure}[!h]
\centering
\includegraphics[width=.57\linewidth]{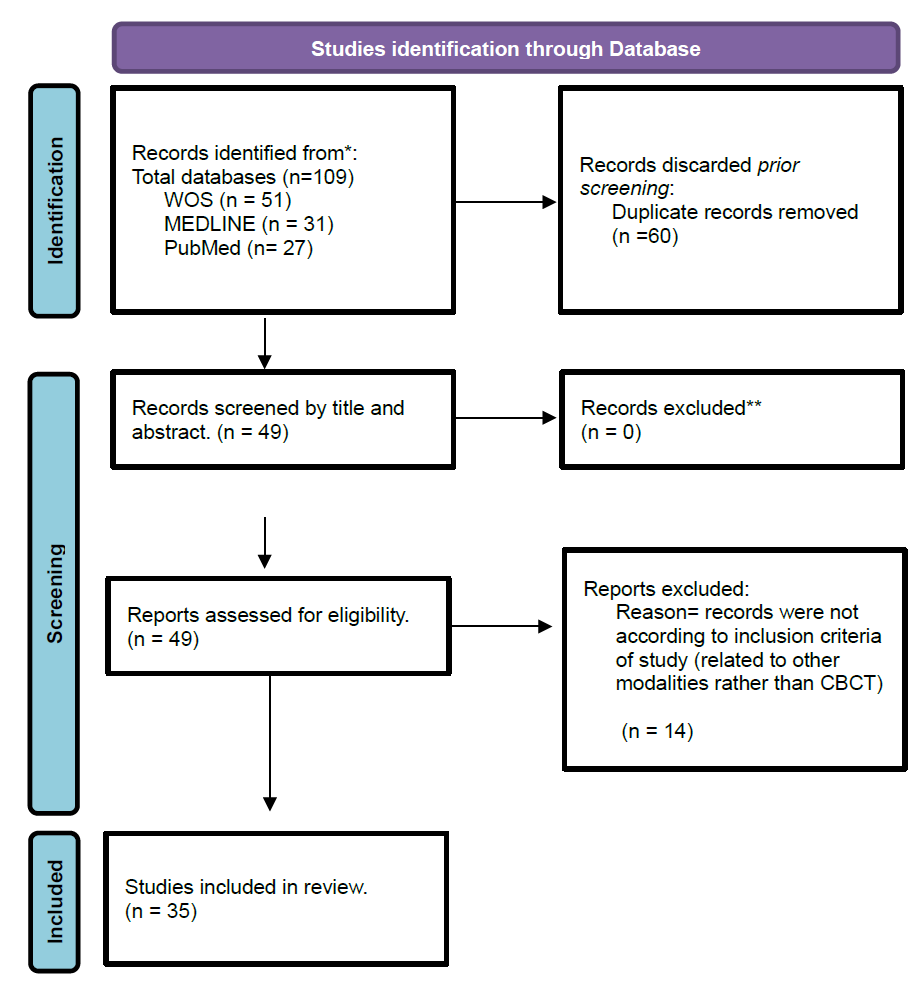}
\caption{PRISMA flowchart depicting the search and study selection process.}
\label{prisma_flowchart}
\end{figure}

The database search identified a total of 109 records: 27 from PubMed, 31 from MEDLINE, and 51 from Web of Science (WOS). After the removal of duplicates, 49 unique records remained. Subsequently, these records were screened according to the established inclusion and exclusion criteria. The first category of excluded articles was for studies that focus on modalities other than CBCT for the generation of synthetic CT (9 papers). These studies mainly investigated imaging modalities such as MRI or EPID, which are not directly related to CBCT. Given that the focus of this research is on the generation of synthetic CT specifically derived from CBCT, these studies were excluded because they did not align with the scope of the review. 
The second exclusion category involved papers with a limited focus on synthetic CT generation from CBCT (5 papers). Although some of these studies discussed CBCT or related aspects of radiotherapy, they did not directly address the generation of synthetic CT from CBCT. As a result, they were excluded because they did not align sufficiently with the primary objective of this study. The remaining 35 records were selected for inclusion in this systematic review. An illustration of the PRISMA screening flow is shown in Fig. \ref{prisma_flowchart}.

In this systematic review, deep learning methods, particularly variants of GANs and CNNs, are predominantly utilized for the generation of sCT from CBCT or other imaging modalities in oncology. These methods offer advantages such as the ability to produce high-quality CT-like images with precise CT numbers, improved dosimetric precision, reduced artefacts, and enhanced anatomical fidelity.

Specifically, techniques such as U-net, CycleGAN, and various GAN variants have been widely adopted due to their effectiveness in generating accurate synthetic CT images across different types of cancer, including head and neck, breast, pelvic, prostate, and thoracic malignancies. Furthermore, transformer-based networks, dosimetry-based clinical workflows, and novel architectures like the Residual Deep Neural Network (RDNN) demonstrate promising results in improving Hounsfield Unit (HU) precision and architectural fidelity in sCT generation.

\begin{figure}[!h]
\centering
\includegraphics[width=\linewidth]{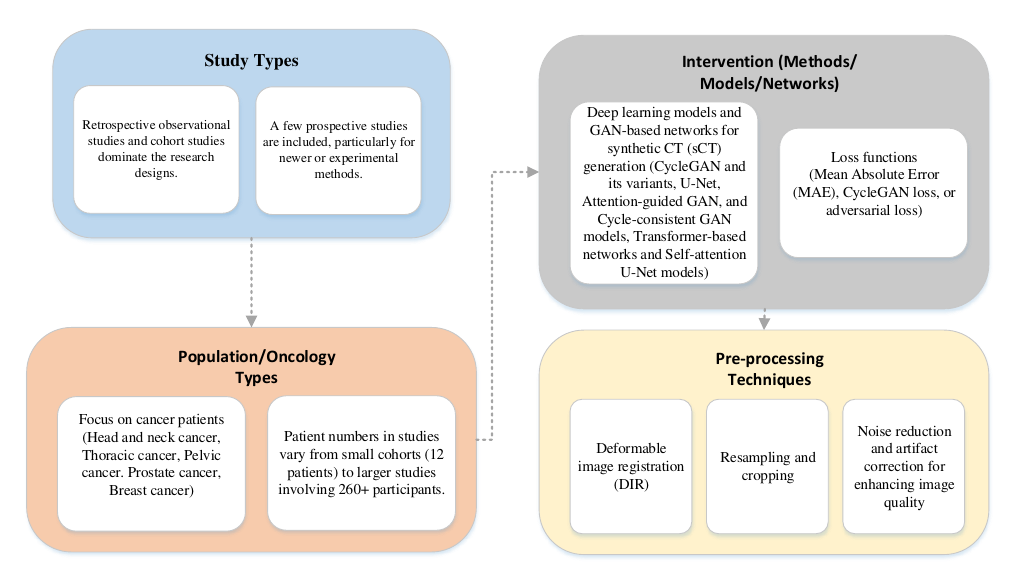}
\caption{Synthesis of the overall findings and key methods involved.}
\label{Synthesis_of_the_Overall_Findings}
\end{figure}

In this context, a synthesis of the general findings and the background of the most widely used models has been presented in Fig. \ref{Synthesis_of_the_Overall_Findings}. 

The study types are retrospective or prospective and incorporate several types of cancer that involve head and neck cancer, thoracic cancer, pelvic cancer, prostate cancer, and breast cancer. The study cohort ranges from 12 to 260+ patients. Most of these studies have been found to reside in some of the commonly used deep learning architectures, including the U-Net, GAN, Cycle-GAN, TransCBCT, and DDPM models, which were detailed earlier in the background section. Pre-processing techniques, loss function, and training methods have also been incorporated into the study.

The list of studies included in this systematic review is given in Table \ref{summary}, together with the data extracted from each study. In addition to the studies above, several proton-based studies have been identified and included in the systematic review. The summary of the findings of these studies is presented in Table \ref{summary_Proton}.

\newpage

\renewcommand{\arraystretch}{1.6}

\begin{table*}[!ht]
\centering
\caption{Summary of findings from all the included articles.}
\label{summary}
\Rotatebox{90}{
\begin{tabular}{p{40pt}p{40pt}p{40pt}p{40pt}p{40pt}p{120pt}p{50pt}p{50pt}p{50pt}p{50pt}}
\hline
\textbf{Article} &	\textbf{Study type} &	\textbf{Population/ Oncology   type} &	\textbf{Intervention (Method of  sCT generation)} &	\textbf{Outcome measures of synthetic CT (MAE, RMSE, PSNR, SSIM)} &	\textbf{Findings} &	\textbf {Pre-Processing Techniques} &	\textbf	{Loss   Function} &	\textbf	{Training Methods} &	\textbf	{Limitations} \\
\hline
Wang et al., 2016 \cite{b32} & 	Retrospective observational study &
Head and neck cancer. &	Group-based or Individual centroids. & N/A &	CBCT-synthesized CTs with individual or group-based centroids yielded dose estimations equivalent to CT-planning dosage for isolated H\&N malignancy. The technique might give a tool for reliable dosage determination utilizing regular CBCT.& Fuzzy c-means classification, centroid calculation &	Mean Absolute Error (MAE)&	Probabilistic classification, voxel-based &	Variability in centroid accuracy between individual and group-based models \\ \hline
Chen et al., 2020 \cite{b2} &	Retrospective observational study & 44 patients with head and neck cancer Validation=7 training= 30, Testing=7 &	Deep learning U-net method &	MAE between rCT and sCT = 18.98 HU, Between rCT and CBCT= 44.38 HU. &	The suggested sCT U-net is capable of producing CT-quality pictures with precise CT numbers using on-treatment CBCT and planned CT. This might enable enhanced CBCT uses for customized planning of treatment.&Deformable registration, resampling &	MAE, hybrid loss &	3D U-Net, paired	&Limited by the small training dataset size  \\ \hline

Liu et al., 2020 \cite{b17} &	Retrospective cohort study &	30 patients treated with SBRT (pancreas) &	Cycle generative adversarial network (cycleGAN) &	MAE between sCT and CT = 56.89 13.84 HU, with comparison to raw CBCT and CT $= 81.06 \pm 15.86$ HU  &	The visual consistency and dosimetric uniformity of the CT versus sCT-based designs verified the dosage calculation precision of the sCT. The CBCT-based sCT technique may improve treatment accuracy and thereby reduce toxicity in the gastrointestinal tract. &Deformable registration, planning CT alignment&	CycleGAN with self-attention, MAE&	Unpaired	&Challenges with patient anatomy change over time. \\ \hline

Grzadziel et al., 2020 \cite{b10}  & Retro prospective study  &	Head \& Torso Free point phantom, specifically model 002H9K. & Deformable image registration &	Discrepancies between the initial plan and the actual delivery varied from 0.0\% to 2.5\%. &	Deformable registration \& synthetic CT can provide effective dose restoration. This study highlights the limitations of the CBCT method, including its limited duration and deformable fusing errors. & Deformable image registration with CBCT	& Not explicitly mentioned	& Deformable registration, 2D	& Limited by CBCT volume length and registration accuracy \\ \hline

Dai et al., 2021 \cite{b4} &	Observational retrospective stud &	1127 patients with breast cancer &	Generative adversarial network &	(MAE) $71.58 \pm 8.78$ HU & This work indicates that deep learning may be used to provide very accurate multi-target segmentation and dosage computation from daily CBCT images. The findings indicate that dose distribution should be evaluated when determining whether to make adjustments during breast cancer radiation to assess the therapeutic impact of mathematical deviations. & Preprocessing of MR and US images	& Weakly-supervised strategy for registration	& 3D V-Net, deformable	& Accuracy is reduced with severe anatomical changes. 
\\ \hline

\end{tabular}
}
\end{table*}

\newpage

\begin{table*}[!ht]
\centering
\Rotatebox{90}{
\begin{tabular}{p{40pt}p{40pt}p{40pt}p{40pt}p{80pt}p{120pt}p{50pt}p{50pt}p{50pt}p{50pt}}
\hline
\textbf{Article} &	\textbf{Study type} &	\textbf{Population/ Oncology   type} &	\textbf{Intervention (Method of  sCT generation)} &	\textbf{Outcome measures of synthetic CT (MAE, RMSE, PSNR, SSIM)} &	\textbf{Findings} &	\textbf {Pre-Processing Techniques} &	\textbf	{Loss Function} &	\textbf	{Training Methods} &	\textbf	{Limitations} \\ \hline

Gao et al., 2021 \cite{b8} &	Retrospective study & 170 patients with thoracic radiotherapy &	Attention-guided generative adversarial networks (AGGAN) & (MAE $= 43.5\pm 6.69$) \par
PSNR= $29.5\pm 2.36$
\par SSIM $= 93.7\pm 3.88$ & The suggested AGGAN was used to produce high-quality sCT images from low-dose thorax CBCT depicts utilizing unpaired CBCT and CT data. Using sCT images in radiation therapy, the dosage distribution may be correctly estimated.&Low-dose CBCT preprocessing, attention-guided &  GAN	Conditional GAN (pix2pix), CycleGAN	& Paired and unpaired	& High computational cost for training models. \\ \hline

Irmak et al., 2021 \cite{b13} &	Retrospective study &	41 patients with head region tumour &	CycleGAN neural network &	Mean Gamma Pass Rate $= 99.0 
\pm 0.4\%$ & The dosimetric findings demonstrate high agreement across sCT, CBCT, and pCT-based computations. A correctly implemented CBCT transformation method can be used as an instrument for quality control in an MR-only radiation process for head victims.& CycleGAN variants, deformable registration &	MAE, RMSE, PSNR, structural similarity index (SSIM)	& Unpaired, 3D	& Error sensitivity in unusual anatomical cases. \\ \hline

(Lerner et al., 2021) \cite{b19} & Prospective &	20 brain malignancy patients &	CNN-based, TFE algorithm &	MAE: 62.2 HU &	CNN-generated sCTs matched CT well for dose planning.& MR-Dixon images, distortion correction	& MAE	& Paired, 3D volumes	 &Small cohort, slice thickness limitation\\ \hline

Oria et al., 2021 \cite{b21} &	Retrospective study & Head and neck cancer patients  &	Deep Convolutional Neural Network (DCNN) method &	Mean Relative Range Error rCTs= 
$-1.2\%$ to $+1.5\%$, and sCTs = $-0.7\%$ - $2.7\%$ & The consistency between observed and generated Relative Proton (RP) fields shows that sCTs are suitable for proton dosage estimations. This finding puts DCNN-generated sCTs towards clinical use in adaptable proton therapy procedures for treatment. The suggested RP QC tool assesses CT number correctness in sCTs that can enable in-vivo range confirmation. & Range probing for quality control&	MSE & 	Deep CNN, unpaired	& QC process complexity in clinical practice  \\ \hline

Rossi and Cerveri, 2021 \cite{b24} &	Retrospective observational study &	58 participants with pelvic region cancer &	Convolutional neural networks (CNNs) &	HU accuracy (62\% vs. 50\%),
\par Peak signal-to-noise ratio (15\% vs. 8\%). 
\par Structural similarity index (2.5\% vs. 1.1\%) &	The research finds that the possibility of using CNN to produce precise synthetic CT via CBCT images is faster and easier to use than conventional methods used in healthcare facilities. &Histogram matching, image-to-image translation &	Supervised: MAE, Unsupervised: Cycle consistency	& Supervised and unsupervised &	Higher artefacts with supervised approach \\ \hline

Xue et al., 2021 \cite{b37} &	Retrospective observational study &	169 patients with nasopharyngeal carcinoma &	CycleGAN, U-Net and Pix2pix models &	Average MAE and RMSE values among sCT by each model and pCT dropped by 15.4 HU or 26.8 HU minimum, whereas average PSNR and SSIM measurements between sCT by various models and pCT increased by 10.6 or 0.05 maximum, correspondingly. &	All sCT produced higher assessment metrics than the initial CBCT, with the CycleGAN modelling outperforming the other two techniques. The dosimetric concordance validated the HU precision and uniform anatomical frameworks of sCT using deep learning algorithms. &CycleGAN, U-Net preprocessing	&MAE, RMSE, SSIM, PSNR	&Paired and unpaired	& Accuracy varies with anatomical differences  \\ \hline

\end{tabular}
}
\end{table*}

\begin{table*}[htbp]
\centering
\Rotatebox{90}{
\begin{tabular}{p{40pt}p{40pt}p{55pt}p{75pt}p{80pt}p{120pt}p{50pt}p{50pt}p{40pt}p{50pt}}
\hline
\textbf{Article} &	\textbf{Study type} &	\textbf{Population/ Oncology   type} &	\textbf{Intervention (Method of  sCT generation)} &	\textbf{Outcome measures of synthetic CT (MAE, RMSE, PSNR, SSIM)} &	\textbf{Findings} &	\textbf {Pre-Processing Techniques} &	\textbf	{Loss Function} &	\textbf	{Training Methods} &	\textbf	{Limitations} \\
\hline

Zhao et al., 2021 \cite{b39} & Retrospective observational study & 40 patients with rectal cancer age range from 38 - 70 (median age = 58). & Cycle-consistent adversarial network (CycleGAN) & MAE was reduced as $135.84 \pm 41.59$ HU comparison to CT and CBCT and $2.99 \pm 12.09$ HU for comparison of sCT and CT & The proposed technique solved a critical obstacle for rectal cancer treatment ART realization using MV CBCT. The resulting sCT facilitates ART according to the patient's real physiology at the therapeutic location. & Deformable registration of MV CBCT, noise reduction	&CycleGAN with MAE, SSIM	& Unpaired, 3D	& High complexity in maintaining HU consistency \\ \hline

Chen et al., 2022 \cite{b3} & Retro prospective study & 91 prostate cancer patients & Transformer-based network (called TransCBCT) & The TransCBCT MAE was $28.8 \pm 16.7$ HU, compared to raw CBCT ($66.5 \pm 13.2$) for CycleGAN ($34.3 \pm 17.3$). & The suggested TransCBCT can successfully create sCT from CBCT. It possesses a chance to increase radiotherapy accuracy.& Transformer-based hierarchical encoder-decoder	& MAE, PSNR, SSIM	& Paired, 3D &	Limited generalizability to other anatomical sites \\ \hline

Deng et al., 2022 \cite{b5} &	Retro prospective observational study &	30 pelvic patients & 	Convolution-consistent (Cycle-RCDC-GAN) &	MAE improved from 28.81 - 18.48, SNU from 0.34 - 0.30, RMSE from 85.66 - 69.50, PSNR from 31.61 - 33.07, SSIM from 0.981 to 0.989. &	Cycle-RCDC-GAN improves CBCT picture quality and has a higher generalization rate than Cycle-GAN, thus encouraging the use of CBCT in radiation therapy. & Dilated convolution, residual connection	& Cycle-consistent GAN	& Paired, 3D	& Generalization challenges across different anatomies \\ \hline

O'Hara et al., 2022 \cite{b20} &	Retro prospective study &	60 patients with head and neck pharyngeal cancer &	Method 1: 
deformable registration of planning CT (pCT) to daily cone-beam CT (dCBCT). Method 2: 
assignment of mass density values to dCBCT. Method 3: 
iteratively removing artifacts and correcting HU of dCBCT. Method 4: 
machine learning model based on Cycle-GAN, trained with paired pCT and CBCT data. & MAE: Method 1 - 59.7 HU, 100.0\%; Method 2 - 164.2 HU, 95.2\%; Method 3 - 75.7 HU, 99.9\%; Method 4 - 79.4 HU, 99.8\% &	All techniques were deemed clinically feasible. Because of its excellent dosimetric precision and fast rate of generation, the method based on machine learning was determined to be the best candidate for clinical deployment. Larger physiological variations among CBCT and pCT, as well as other structural locations, necessitate more examination. & Deformable registration, artefact correction &	MSE	& Paired, 3D &	Time-consuming data preparation process  \\ \hline

Wang et al., 2022 \cite{b34} &	Retrospective study  &	68 patients under radiotherapy after surgery of breast-conserving  &	Comparison of 3 deep-learning methods for sCT synthesis from CBCT, Cycle generative adversarial network (CycleGAN), U-Net, and pix2pix. &	The U-Net model had the lowest possible mean absolute error (MAE) 62.53 × 9.14 HU within the body. & The U-Net model can create sCT pictures with greater image consistency and dosimetric precision than the pix2pix or CycleGAN models. The technique might be utilized to provide precise dose estimation for cancer treatment with adaptive radiotherapy using CBCT. & Pre-processing of breast CBCT data	& MAE, SSIM	& Paired (U-Net, CycleGAN)	& Variability in dose calculation due to anatomy changes \\ \hline

Wu et al., 2022 \cite{b35} &	Research study  &	153 patients with prostate cancer (33 testing patients 33, and training patients 120). &	Multiresolution residual deep neural network (RDNN) &	MAE between CBCT and pCT was 352.56 HU. MAE between sCT and pCT was 52.18 HU &	The suggested multiresolution RDNN generates sCT images with improved HU precision and architectural fidelity.& Multiresolution strategy, noise suppression	& Mean Absolute Error(MAE), SSIM & 	Paired, 3D	& Overfitting risk in smaller datasets \\ \hline

\end{tabular}
}
\end{table*}

\begin{table*}[htbp]
\centering
\Rotatebox{90}{
\begin{tabular}{p{40pt}p{40pt}p{50pt}p{75pt}p{80pt}p{120pt}p{50pt}p{50pt}p{40pt}p{50pt}}
\hline
\textbf{Article} &	\textbf{Study type} &	\textbf{Population/ Oncology   type} &	\textbf{Intervention (Method of  sCT generation)} &	\textbf{Outcome measures of synthetic CT (MAE, RMSE, PSNR and SSIM)} &	\textbf{Findings} &	\textbf {Pre-Processing Techniques} &	\textbf	{Loss Function} &	\textbf	{Training Methods} &	\textbf	{Limitations} \\
\hline

Allen et al., 2023 \cite{b1} &	Retrospective &	12 Head \& Neck cancer patients &	Dosimetry-based clinical workflow &	N/A &	Synthetic CTs are a significant improvement for the adaptive radiotherapy process, and synthetic CT dosage estimations may be utilized successfully with the existing technique of visually reviewing the image overlay of the planned CT and CBCT to judge the relevance of anatomical alteration.& Deformable image registration of pCT to CBCT &	MSE	& Unpaired, 3D & Dosimetric uncertainties for anatomical changes \\ \hline

Gao et al., 2023 \cite{b7} &	Retrospective study &	260 cancer patients, including 140 abdominal and 120 thoracic CT scans &	Streaking artefact reduction network (SARN) combined with cycleGAN &	SARN+cycAT achieved 42.5 HU MAE within the thorax, whereas SARN+cycleGAN achieved 32.0 HU MAE across the abdomen.
&	The sCT produced by U-Net has much worse anatomical structural accuracy than the other algorithms. The pulmonary and abdominal sCT images produced by SARN+ cycle GAN demonstrated the best dose calculation precision, with a gamma rate of passage (2 mm/\%) of 98.1 percent 96.9\%, correspondingly.&Streaking artifact reduction with SARN & SARN + CycleGAN	& Paired, 3D	& Limited clinical validation for larger anatomical changes \\ \hline

Liu et al., 2023 \cite{b16} &	Retrospective study &	230 Chinese female patients with cervical cancer with a mean age of 58 years  &	Encoder–decoder architecture with residual learning and skip connections &	The mean absolute error (MAE) among the model's synthetic CT pictures and planned CT were 10.93 HU, however, the CBCT scans were 50.02 HU.
\par
SSIM increased from 0.76 to 0.90.
\par
PSNR increased from 27.79 dB to 33.91 dB &	The method can produce CT images with higher picture quality and more precise HU measurements. The generated CT images effectively conserved tissue borders, which is critical for subsequent adaptive radiation jobs.&Hierarchical training, residual learning	& MAE, PSNR, SSIM	& Paired, 3D	& Limited to cervical cancer datasets \\ \hline

Suwanraksa et al., 2023 \cite{b27} &	Retrospective study
&	146 patients diagnosed with head and neck cancer &	Generative adversarial networks (GANs) with registration network (RegNet) &	MAE (from 40.46 to 37.21)
RMSE (from 119.45 to 108.86) 
\par
PSNR (from 28.67 to 29.55) 
\par
SSIM (from 0.8630 to 0.8791) &	The model-generated sCT had fewer artefacts and preserved anatomical data, similar to CBCT.& Registration-based correction, noise reduction&	GAN with RegNet	&Supervised with noisy labels&	Challenges with accurate registration of anatomy  \\ \hline

Szmul et al., 2023 \cite{b28} &	Research study & N/A & cycle-consistent Generative Adversarial Networks (cycleGANs) & CBCT vs synthetic CT, MAE ($49.8 \pm 10.9$ vs $55.0 \pm 16.6$) NCC ($0.97 \pm 0.02$ vs $0.97 \pm 0.02$) RSME ($88.6 \pm 24.9$ vs $89.8 \pm 23.8$) &	Results suggest that our modifications to the cycleGAN framework enhanced the quality and structural consistency of the synthesized CT images produced.& 2D slice selection, weakly paired data approach	&Modified CycleGAN loss with structural consistency	& CycleGAN, 2D	&Limited to small pediatric datasets \\ \hline

Kong et al., 2023 \cite{b14} & Retrospective cohort study. & 35 patients with head and neck cancer & An enhancement technique with random style (Shuffle Remap) with a deep learning generator & The MAE within the body were $(44.7 \pm 2.16$ HU vs. $36.7 \pm 131.4$ HU) and $(64.9 \pm 123.7$ HU vs. $58.2 \pm 152.8$ HU) for the CBCT, SCT, and MRI-SCT accordingly. & The method demonstrated great accuracy in dosimetric measurements and outstanding IMRT QA verification outcomes. The proposed SmGAN, unlike previous synthetic CT creation methods, only requires a single-modal picture for training. This is a significant development in the industry and has potential for several uses in medicine.&Spatial transformations, style enhancement	&Mean Absolute Error (MAE)	& Paired, 3D (SmGAN)	& Requires extensive training time for model convergence \\ \hline

\end{tabular}
}
\end{table*}

\begin{table*}[htbp]
\centering
\Rotatebox{90}{
\begin{tabular}{p{40pt}p{40pt}p{50pt}p{75pt}p{80pt}p{120pt}p{50pt}p{50pt}p{40pt}p{50pt}}
\hline
\textbf{Article} &	\textbf{Study type} &	\textbf{Population/ Oncology   type} &	\textbf{Intervention (Method of  sCT generation)} &	\textbf{Outcome measures of synthetic CT (MAE, RMSE, PSNR and SSIM)} &	\textbf{Findings} &	\textbf {Pre-Processing Techniques} &	\textbf	{Loss Function} &	\textbf	{Training Methods} &	\textbf	{Limitations} \\
\hline

Wan et al., 2023 \cite{b31} &	Retrospective cohort study. &	74 patients with Prostate cancer  &	Cycle adversarial generative network (CycleGAN) &	MAE within the region of interest like bladder, muscle, body, and fat among the pCT and sCT are 41.2, 26.5, 29.0, and 25.1 HU, respectively. &	This study uses deep learning algorithms to create a dose-accurate adaptive procedure for prostate cancer individuals, as well as deploy ART that responds to bladder dosage. It is worth noting that the specific replanning parameter for whether ART is required can be adjusted to reflect the preferences of individual centres determined by their knowledge and findings.& Deformable image registration, U-net for contouring	& CycleGAN loss, MAE	& Paired, 3D	& Dependence on accurate bladder delineation for ART  \\ \hline

Wang et al., 2023 \cite{b33} & Retrospective cohort study. & 150 patients undergoing radiotherapy for esophageal cancer. 120 training patients and 30 testing patients & Registration generative adversarial networks (RegGAN) & AE CBCT vs sCT: $80.1 \pm 9.1$ vs $43.7 \pm 4.8$; CBCT vs RMSE sCT: $124.2 \pm 21.8$ vs $67.2 \pm 12.4$, and CBCT vs PSNR sCT: $21.3 \pm 4.2$ vs $27.9 \pm 5.6$ & The designed deep-learning RegGAN system appears interesting for efficiently generating high-quality sCT images derived from stand-alone thoracic CBCT pictures, potentially supporting CBCT-based cancer of the esophageal adaptive radiotherapy.&Registration of planning CT to CBCT	& MAE, Root Mean Square Error (RMSE), PSNR &	2.5D RegGAN, unsupervised	& Limited to esophageal cancer patients  \\\hline

Wynne et al., 2023 \cite{b36} &	Research study &	79 patients receiving prostate cancer proton therapy &	Contrastive unpaired translation (CUT), CycleGAN and GAN &	Mean Absolute Error (HU), cycleGAN (47.12), CBCT (22.42), and CUT (19.52)
\par RMSE: cycleGAN 149.37 (72.81), CBCT 65.00 (11.97), CUT 74.312 (11.28) &	The technique against cycleGAN utilizing structural similarity mean absolute error, index, Frèchet Inception Distance and root mean squared error, demonstrates that CUT outperformed cycleGAN while utilizing fewer resources as well as time. The researched approach advances the practicality of online adaptive radiation beyond the current state-of-the-art.& Contrastive unpaired translation (CUT)	& Fréchet Inception Distance, MAE, SSIM	& Unpaired, 2D	Struggles with preserving subtle anatomical details  \\ \hline

Yoganathan et al., 2023 \cite{b38} & Research study  &	93 patients with Head and neck cancer &	'Self-attention-residual-U-Net' (Res U-Net) &	MAE: $46\pm 7$ HU, PSNR: $66.6\pm 2.0$ dB, SSIM: $0.81\pm 0.04$ &	Res U-Net improved CT number image quality and accuracy of sCT, outperforming U-Net in sCT production from CBCT. This approach shows potential in producing accurate sCT for the Head and neck for adaptive radiotherapy.& Self-attention mechanism, residual learning	& MAE, PSNR, SSIM	& 2.5D ResUNet, adversarial	& Requires substantial computational resources \\ \hline

Liang et al., 2024 \cite{b15} & Research study & 58 patients having pelvic lesions & Generative adversarial network (GAN) & MAE $(13.02 \pm 4.40$ HU), PSNR $(37.53 \pm 3.06$ dB), RSME $(56.59 \pm 20.58$ HU), SSIM $(93.43 \pm 1.23\%)$ & This study introduces innovative task-specific loss coefficients into a GAN-based network to generate high-quality sCT pictures. Furthermore, the model excels at removing artefacts and suppressing noise. The dosimetric data demonstrate the possibility of inclusion into the clinical ART practice.& Gradient and style loss, task-specific regularization	& Gradient loss, style loss	& GAN-based, paired	& Complexity increases with additional loss functions \\ \hline

Peng et al., 2024 \cite{b23} &	Research study  &	Head and neck cancer &	Conditional denoising diffusion probabilistic model (DDPM) & 	Brain patient study: sCT MAE: 25.99 HU, PSNR: 30.49 dB, NCC: 0.99; H\&N patient study: sCT MAE: 32.56 HU, PSNR: 27.65 dB, NCC: 0.98. &	The proposed conditional DDPM approach can produce sCT from CBCT with precise HU values and minimal artefacts, enabling accurate CBCT-based dose calculation and tissue delineation for digital adaptive radiotherapy (ART).& Conditional diffusion model, residual blocks	& Diffusion loss, MAE, PSNR	&Paired, conditional diffusion model&	High computational cost due to diffusion processes  \\ \hline

\end{tabular}
}
\end{table*}

\begin{table*}[htbp]
\centering
\caption{Summary of the Findings of Proton-Based articles.}
\label{summary_Proton}
\Rotatebox{90}{
\begin{tabular}{p{40pt}p{40pt}p{40pt}p{40pt}p{45pt}p{40pt}p{40pt}p{30pt}p{30pt}p{40pt}}
\hline
\textbf{Article} & \textbf{Study Type} & \textbf{Population/ Oncology Type} & \textbf{Intervention (Method/Model/ Network  of sCT generation)} & \textbf{Outcome Measures of sCT (MAE, RMSE, PSNR, SSIM)} & \textbf{Findings} & \textbf{Pre-Processing Techniques} & \textbf{Loss   Function} &\textbf{Training Methods}&\textbf{Limitations} \\ \hline

(Thummerer, Zaffino, et al., 2020) \cite{b48} & Retrospective & 33 head \& neck cancer patients & DCNN (Unet); Loss: MAE & MAE: 36.3 HU, Gamma: 99.43\% & DCNN provided highest image quality & DIR, cropping, segmentation & MAE&	Paired, 3D volumes &Head \& neck only, AIC method under development \\ \hline

(Thummerer, de Jong, et al., 2020) \cite{b57} & Retrospective & 27 head \& neck cancer patients & DCNN (Unet); Loss: MAE & MAE: 40.2 HU (CBCT), 65.4 HU (MR) & CBCT-based sCTs had better image quality than MR-based & DIR, cropping, segmentation & MAE&	Paired, 3D volumes  & Limited dataset, registration errors for MR \\ \hline

(Thummerer et al., 2021) \cite{b52} & Retrospective & 33 thoracic cancer patients & DCNN; Loss: MAE, L1 regularization & MAE: 30.7 HU, PSNR: 31.2 dB, SSIM: 0.941 & DCNN-based sCTs accurate for adaptive proton therapy & Rigid registration, DIR, cropping & MAE, L1 &	Paired, 3D volumes & Limited cohort, CBCT quality sensitivity \\ \hline

(Pang et al., 2023) \cite{b43} & Retrospective & 75 nasopharynx cancer patients & cGAN; Loss: Adversarial, L1 & MAE: 16.39 HU, PSNR: 28.98 dB, SSIM: 0.9524 & cGAN outperformed others in dose accuracy & Rigid registration, DIR, cropping& Adversarial, L1	& Unpaired, 3D volumes & Limited to NPC patients, requires validation for other regions \\ \hline

(Taasti et al., 2023) \cite{b41} & Retrospective & 42 lung cancer patients & CBCT \& DIR-based & Gamma pass: 93\%, False negatives: 5\% & Both sCT methods are suitable for reducing CT acquisition & CBCT correction, deformable registration & MAE&	Paired, 3D volumes & Limited to lung cancer, small cohort \\ \hline

(Perez Moreno et al., 2023) \cite{b26} & Retrospective & Exvivo phantom using pork and cow tissues & Synthetic CT generated using the "Virtual CT" algorithm of RayStation 12B &	Estimated range uncertainty for CT: 1.6\%, for CBCT: 2.9\% &	CBCT has higher range uncertainty (2.9\%) than CT (1.6\%), but it's suitable for online adaptive therapy & - & MAE &	Paired, 2D slices &	Uncertainties are high in air-tissue interfaces and dense bone areas \\ \hline

\end{tabular}
}
\end{table*}

\subsection {Quantitative Results}
\begin{figure}[!h]
\centering
\includegraphics[width=.5\linewidth]{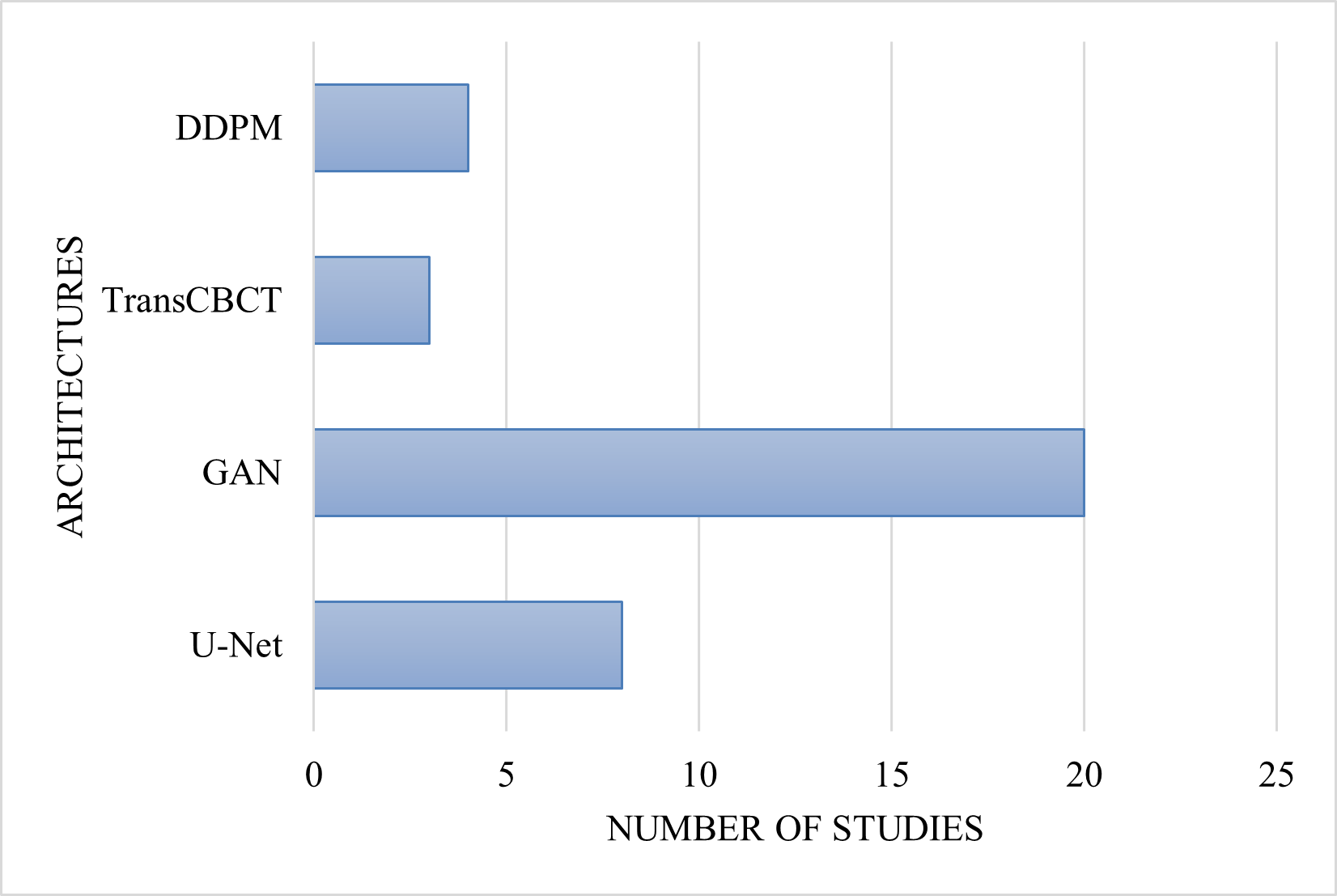}
\caption{Distribution of articles based on the primary architectures identified.}
\label{the_primary_architectures_identified}
\end{figure}

To analyze the distribution of articles under several architectures and outcomes, a statistical analysis has been carried out. Initially, the distribution of DL architectures has been analyzed in their primary categories and their variants. The four main categories include U-Net, GAN, TransCBCT, and DDPM. These categories have been selected because of their prevalence in the reviewed literature with their adaptation in sCT image generation. The distribution of the 35 studies published from 2016 to 2024 based on these primary architectures and subsequent variants is presented in Fig. \ref{the_primary_architectures_identified}.

The focus has been kept on DL architectures applied to CBCT-to-CT image synthesis. The most widely used architectures include U-Net, GAN, and TransCBCT. Recent studies indicate a shift toward more complex models like DDPM and transformer-based networks. These models are better suited for handling intricate imaging artifacts and enhancing the feasibility of clinical application. Emerging diffusion models, such as DDPM, show considerable promise especially for the generalizability power on unseen data, which is an aspect currently under extensive investigation and for future research.

\begin{figure}[!h]
\centering
\includegraphics[width=.5\linewidth]{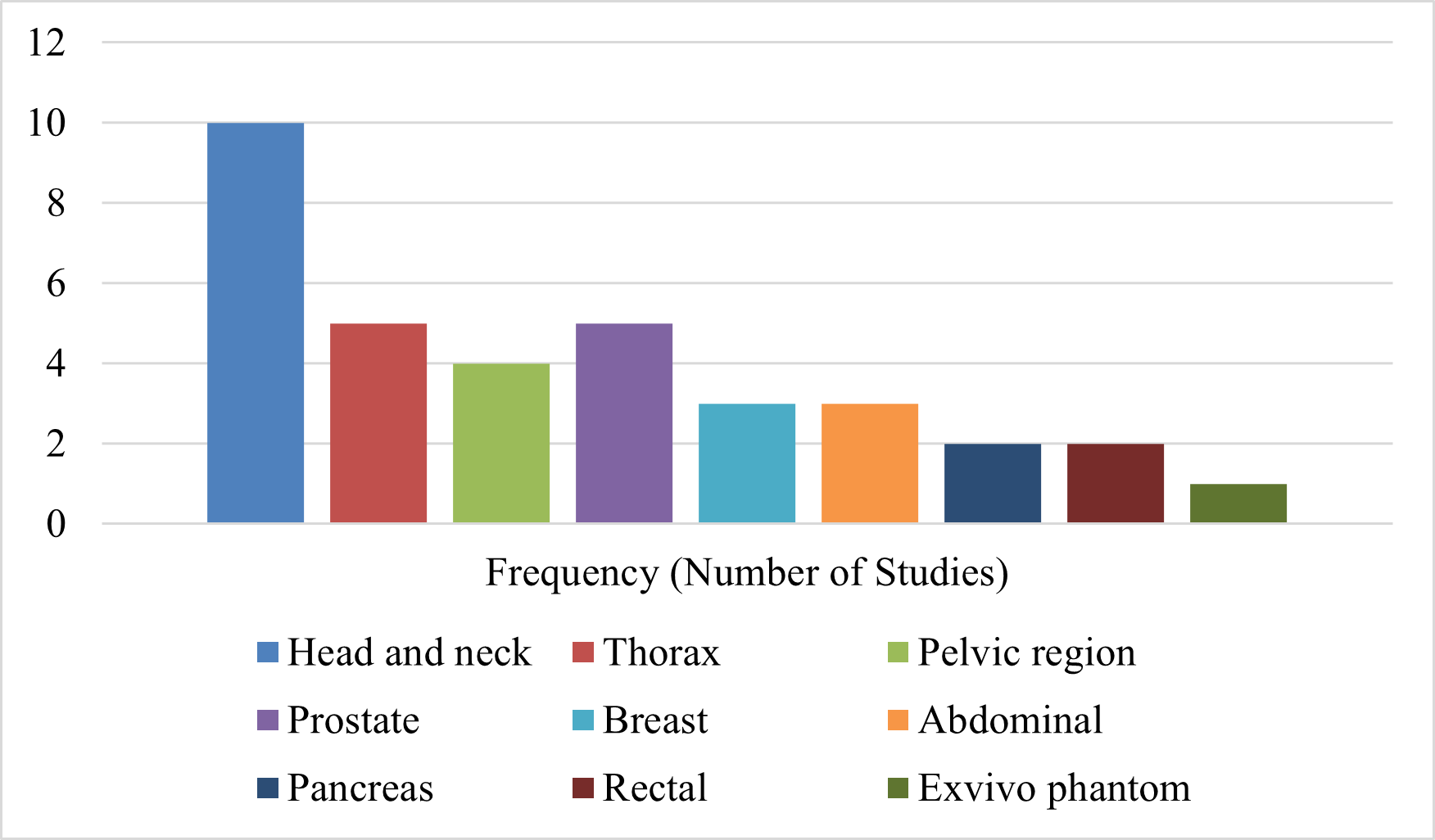}
\caption{Distribution of articles based on the anatomical regions.}
\label{Anatomical_Regions}
\end{figure}

The second analysis carried out is based on anatomical regions. The distribution of the anatomical regions includes head and neck, thorax, pelvic region, prostate, breast, abdomen, pancreas, rectal, and ex vivo phantom studies. Head and neck cancer has been found to be the focal point of the investigation along with thoracic radiotherapy, as shown in Fig. \ref{Anatomical_Regions}.

\begin{figure}[!h]
\centering
\includegraphics[width=.5\linewidth]{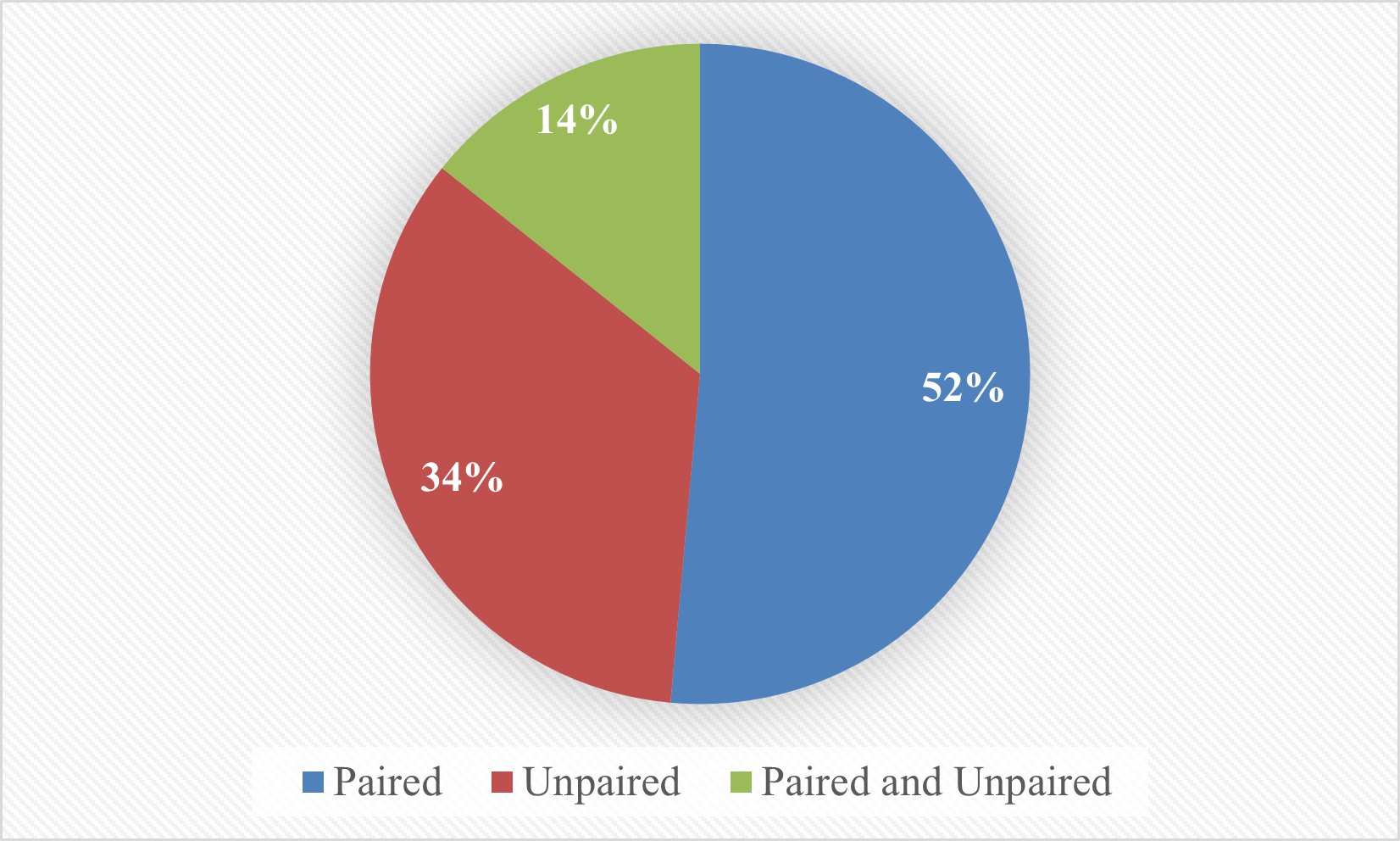}
\caption{Distribution of articles based on the training methods.}
\label{Articles_based_on_the_Training_Methods}
\end{figure}

In the context of training methods being used, the distribution has been presented in Fig. \ref{Articles_based_on_the_Training_Methods}. The training methods have broadly been classified into paired, unpaired, and a combination of both. Most studies have found that they make use of paired training methods in which all slices are kept intact as 3D volumes. Several studies also make use of unpaired methods to overcome some of the challenges in data pairing. 

\begin{figure}[!h]
\centering
\includegraphics[width=.5\linewidth]{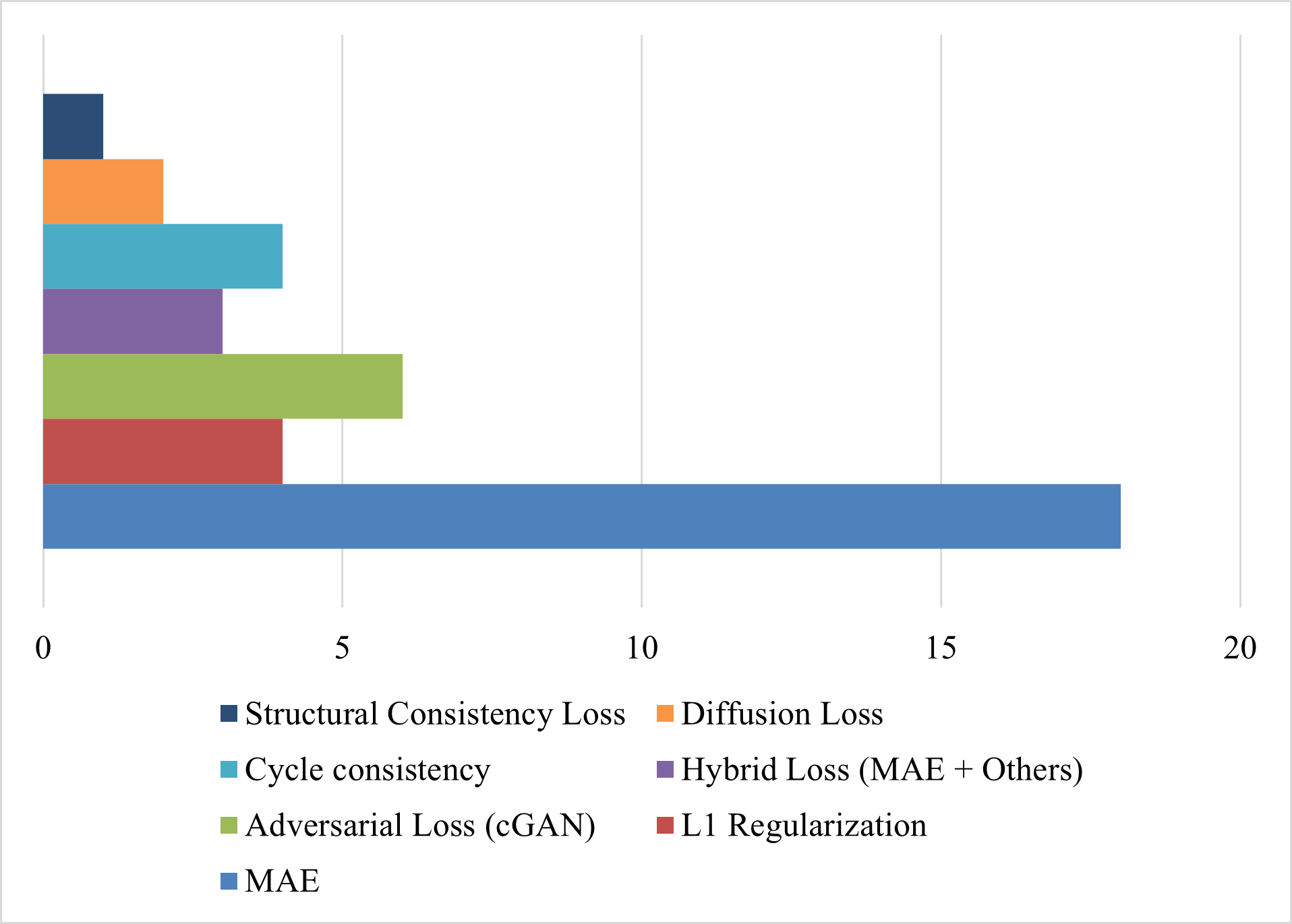}
\caption{Distribution of articles based on the loss functions.}\label{Articles_based_on_the_Loss_Functions}
\end{figure}

The loss function is an important consideration in the development of deep learning models. Several types of loss functions have been adopted as part of sCT image generation. Some of the most commonly employed methods are MAE (individual or combined with $L_1$ regularization, etc.).  The graphs are presented in Fig. \ref{Articles_based_on_the_Loss_Functions} and demonstrate the distribution of loss functions in the included articles.

\begin{figure}[!h]
\centering
\includegraphics[width=.5\linewidth]{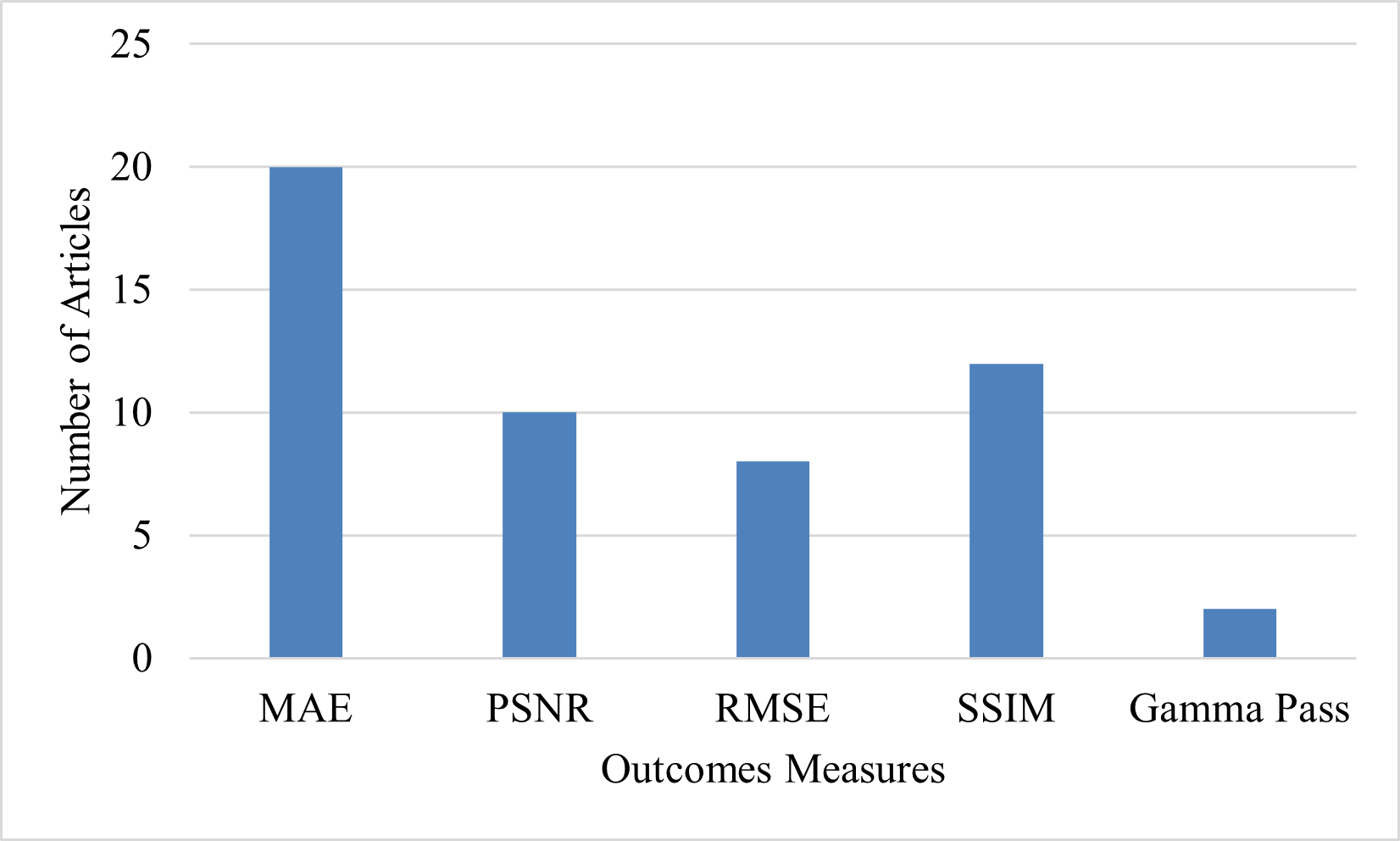}
\caption{Distribution of articles based on the performance outcomes.}\label{Articles_based_on_the_Performance_Outcomes}
\end{figure}

Finally, the outcomes used in the articles have been explored. MAE has been found to be the most frequently used outcome measure in addition to PSNR, RMSE, and SSIM. The analysis indicated that the studies largely employ quantitative image similarity to validate the accuracy of the models. However, there is a gradual inclination towards dosimetric measures such as Gamma Pass for clinical validation. The distribution is presented in Fig. \ref{Articles_based_on_the_Performance_Outcomes}.

The quantitative analysis of the 35 studies included in the study has shown some of the key trends and gaps in the research on the generation of sCT images for adaptive radiotherapy. Most studies have been found to rely on the use of U-Net and GAN (including CycleGAN) architectures due to their ability to handle complex medical imaging tasks. Some of the less commonly used models, such as TransCBCT and DDPM, are beginning to emerge; however, they have yet to be thoroughly explored. The anatomical regions studied predominantly include the head, neck, and thoracic areas. These areas have been prioritized due to the clinical importance of radiotherapy. The other anatomical regions, such as the pancreas and rectal cancer, remain relatively unexplored. Thus, further studies are imperative in these regions. In the context of training methods, the paired datasets are more commonly used. However, the presence of unpaired datasets demands and ongoing effort to overcome the challenges of data pairing. The use of MAE as a loss function and performance outcome prevails. This needs to be considered in the future to employ and experiment with the other loss functions. Additionally, the development of custom loss functions according to the data set needs is less commonly explored.

\subsection{Registration and Reporting} 
This systematic review was conducted according to a protocol that was previously registered with the International Prospective Register of Systematic Reviews (PROSPERO), under the registration number CRD42024540973. The findings were reported according to the PRISMA guidelines.

\section{Dosimetry Accuracy and Quality of Imaging} 
Overall, the dosimetric values reported in the various studies indicate the effectiveness of synthetic CT (sCT) generation methods to accurately estimate dose distributions for radiation therapy planning. MAE, RMSE, PSNR, SSIM, and other metrics were used to assess the fidelity of the sCT images compared to the gold standard planning CTs (pCTs) and to evaluate the precision of dose calculations. In all studies, the reported MAE values ranged from approximately 10 HU to 135 HU, with lower values indicating a better agreement between sCT and pCT. Similarly, RMSE values ranged from around 18 HU to 124 HU, with smaller values indicating a more accurate dose estimation. PSNR values, which measure image quality, ranged from approximately 21 to 67 dB, with higher values indicating better image fidelity. SSIM values, which assess structural similarity between images, ranged from around 0.79 to 0.99, with values closer to 1 indicating greater similarity. In addition, some studies reported specific dosimetric parameters such as gamma index analysis, which assesses the agreement between planned and delivered doses, and dose distribution evaluations for various anatomical regions. In general, the reported dosimetric values demonstrate the potential of sCT generation methods, particularly those utilizing deep learning techniques, to produce accurate and clinically useful images for radiation therapy planning, thus improving treatment precision and patient outcomes.

\begin{figure*}[!h]
\begin{center}
\includegraphics[width=.7\linewidth]{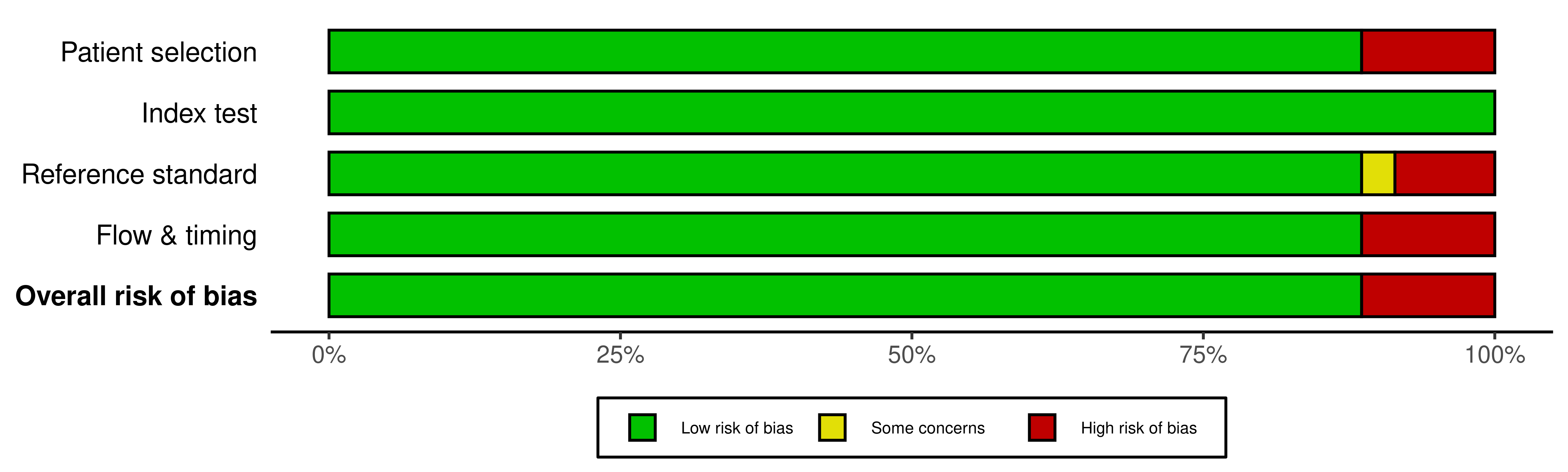}
\caption{The risk of bias assessment results for all the included studies were conducted using QUADAS 2.}\label{Quadas}
\end{center}
\end{figure*}

\section{Risk of Bias Assessment} 
In the assessment of included studies, it was found that all the studies showed a low risk of bias in various domains, as shown in Fig. \ref{Quadas}. In general, only two studies were classified as having high risk in this analysis \cite{b10}, \cite{b26}, \cite{b57}, \cite{b32}. Of the 35 studies reviewed, four demonstrated a high risk of patient selection bias \cite{b10}, \cite{b26}, \cite{b28}, \cite{b57}; another three exhibited a high risk related to the reference standard \cite{b32}, \cite{b10}, \cite{b41}; a study raised some concerns about the reference standard \cite{b1} and four studies raised a high risk related to flow and timing \cite{b19}, \cite{b26}, \cite{b57}, \cite{b52}. This indicates a high level of methodological rigor and reliability in the findings reported within the systematic review. As such, the results presented here can be considered robust and trustworthy, suggesting that the synthesized evidence has significant value to inform both clinical practice and future research endeavors in the domain of synthetic CT generation for oncology applications.

\section{Discussions} 
The findings of the systematic review suggest that deep learning methods have shown a promising role in the generation of sCT images from CBCT images. The findings have been particularly analyzed in the context of oncology and radiation therapy planning. Some of the predominant techniques have been included in the articles that include architectures such as U-Net, GAN, CycleGAN, TransCBCT, and DDPM models. The models have depicted varying levels of effectiveness, and each of them has offered distinct advantages and challenges. The performance of sCT generation is largely dependent on the deep learning architectures employed. For example, U-Net, which has a simplified encoder-decoder architecture, has been commonly used, and it retained high-resolution image details. This results in better MAE and SSIM values as analyzed in the reviewed articles. The performance of the U-Net becomes limited when complex anatomical structures are handled. For example, in the presence of artifacts in the images, advanced models such as CycleGAN and TransCBCT outperform U-Net’s performance. 

GANs and variants such as CycleGANs have been used prominently in the generation of high-quality sCT images to offer improved dosimetric accuracy. The models are particularly effective in unpaired image-to-image translation tasks. This becomes critical in the nonavailability of the paired datasets. The adversarial loss mechanism embedded into such architectures helps in offering improved performance in complex anatomies such as head and neck regions. The complexity of GANs also increases the computational requirements and training time, limiting their application in clinical settings. 

Emerging models such as TransCBCT and DDPM have shown their potential to improve image quality. The TransCBCT has been found to have better HU precision compared to the traditional models that are being used. The transformed-based models as well as the diffusion models thus hold promising approaches in the research. However, complexity and computational cost limit their applicability on larger scales. 

\textbf {Advantages and Challenges Across Oncology Types}: the articles reviewed cover various types of cancer, including head and neck, thoracic, pelvic, and prostate cancers, etc. Among these, the head and neck regions are relatively more targeted, and deep learning models such as U-Nets and GANs have been applied largely to them. The complexities of anatomy pose several challenges, including maintaining greater accuracy in several types of tissues and addressing the anatomical changes over time. For thoracic and abdominal cancers, GANs performed comparatively well in generating sCT images with high PSNT and SSIM. However, the regions also present unique challenges due to respiratory motion. Such motions can introduce artifacts in the imaging process. The pancreatic and rectal regions are relatively less explored. Thus, more research in these regions is imperative to validate sCT generation methods. 

In prostate cancer, CycleGAN-based models have shown effectiveness in terms of dose calculations and the generation of anatomical precision. Yet some of the challenges were identified as part of studies including an accurate delineation of bladder and rectum structured. These can affect the overall dosimetric performance of the model. 

\textbf {Limitations and Bias}: although the studies presented above offered promising results, there are some limitations. One of the issues presented was the limited size and variability of the datasets used in the training of the models. Many studies have resided on the small cohorts and the use of paired datasets. These can result in bias and may fail to offer generalizations in real-world clinical settings. High computational cost and training times of the advanced models like GANs and DDPM also make it difficult to be embedded into clinical practices. 

Another key limitation observed is the lack of standardization in terms of evaluation metrics. MAE, PSNR, SSIM, and RMSE are commonly used, but the use of dosimetry measures such as Gamma Pass is less commonly used. This is essential to validate the clinical utility of the sCT images. In addition, most studies are based on the use of retrospective data, which can cause selection bias. The datasets may fail to capture the diversity of the patient population in clinical practices. 

\textbf {Answers to Research Question}: the studies primarily targeted the question of whether the integration of deep learning methodologies for the generation of sCT images from CBCT data enhances the accuracy and effectiveness of radiation therapy planning in oncology and potentially reduces exposure. 

The answer to this question is found to be positive. The integration of deep learning methods can significantly increase the accuracy and effectiveness of radiation therapy planning. The methods reviewed as part of the included articles, such as the U-Net, CycleGAN, and GAN variants, have shown potential to demonstrate improved sCT image generation compared to conventional methods. The methods help reduce artifacts in the images and improve the dosimetric accuracy for the planning of radiation therapy. The use of sCT from CBCT images also has the potential to reduce the exposure to patients in terms of repeated CT scans. As models can generate CT-quality images from the low-quality CBCT data, exposure to radiation is reduced. However, some of the challenges including the limitations of the dataset sizes can increase computational codes and remain a challenge in their applicability to clinical settings.

\section{Conclusion} 
In conclusion, the synthesis of synthetic CT (sCT) images from cone beam CT (CBCT) data through various deep learning methodologies represents a significant advance in radiation oncology. Across the reviewed studies, diverse techniques such as CycleGAN, U-Net, Pix2pix, and attention-guided generative adversarial networks (GANs) have been employed, each demonstrating promising results in accurately estimating dose distributions for radiation therapy planning. The reported dosimetric values, including MAE, RMSE, PSNR, and SSIM, highlight the efficacy of sCT generation methods in producing images with fidelity comparable to gold standard planning CTs (pCTs). These findings underscore the potential of sCT-based approaches to improve precision of treatment and patient outcomes in various cancer scenarios, including head and neck cancer, pancreatic cancer, breast cancer, and thoracic malignancies. Furthermore, the ability of deep learning algorithms to generate high-quality sCT images from CBCT data offers a promising avenue for personalized treatment planning and adaptive radiotherapy. In the future, continued research efforts, standardization of evaluation protocols, and integration into clinical workflows are essential to realize the full potential of sCT technology in improving the delivery and outcomes of cancer treatment. By addressing these challenges, sCT-based approaches have the potential to revolutionize radiation therapy planning and contribute to improved patient care in the field of oncology.

\section*{Acknowledgement}

This work did not involve human subjects or animals in its research.

All authors declare that they have no known conflicts of interest in terms of competing financial interests or personal relationships that could have an influence or are relevant to the work reported in this paper. A. Perelli acknowledges the support of the Royal Academy of Engineering under the RAEng / Leverhulme Trust Research Fellowships program (award LTRF-2324-20-160).

\bibliography{refs}

\end{document}


%
\title{Supplementary Material for the article\\"Synthetic CT image generation from CBCT:\\A Systematic Review"}
\author{Alzahra~Altalib,~Scott~McGregor,~Chunhui~Li,~Alessandro~Perelli 
\thanks{The work did not involve human subjects or animals in its research. A. Altalib, C. Li and A. Perelli  are with the Centre of Medical Engingeering and technology, University of Dundee, DD1 4HN, UK, e-mail: \{2600129, c.li, aperelli001\}@dundee.ac.uk. A. Perelli is supported by the Royal Academy of Engineering under the RAEng / Leverhulme Trust Research Fellowships (award LTRF-2324-20-160). S. McGregor is with the Library \& Learning Centre, University of Dundee, DD1 4HN, UK, e-mail: S.V.Mcgregor@dundee.ac.uk.}
\thanks{Manuscript received December 23, 2024.}}

\markboth{IEEE~Transactions~on~Radiation~and~Plasma~Medical~Sciences,~Vol.~, No.~, ~2025}%
{A. Altalib \MakeLowercase{\textit{et al.}}: Synthetic high-quality CT Image generation from low-quality Cone-Beam CT}

\maketitle

\section{Databases and Keywords used in the systematic review}

\renewcommand{\arraystretch}{1.2}

\begin{table*}[!h]
\centering
\begin{tabular}{p{100pt}p{250pt}}
\hline
\textbf{Database} &	\textbf{Keywords}  \\ \hline

\textbf{WOS} & LA=(English) AND PY=(2014-2024) AND TS= (("Synthetic CT”) AND (“CBCT”) AND (“Dose calculation")) \\

\textbf{MEDLINE} & LA=(English) AND DOP=(2014-2024) AND TS= (("Synthetic CT”) AND (“CBCT”) AND (“Dose calculation")) \\

\textbf{PUBMED} & ("Synthetic CT"[Title/Abstract] OR Synthetic CT[All Fields]) AND ("CBCT"[Title/Abstract] OR CBCT[All Fields]) AND ("Dose calculation"[Title/Abstract] OR "Dose calculation"[All Fields]) AND English[Language] AND ("2014/01/01"[Date - Publication] : "2024/12/31"[Date - Publication]) \\
\hline
\end{tabular}

\end{table*}

\section{Parameters' Risk for each evaluated study}

\begin{table*}[!h]
\centering
\begin{tabular}{p{100pt}p{70pt}p{50pt}p{80pt}p{70pt}p{50pt}}
\hline

\textbf{Study} & \textbf{D1 Patient selection} & \textbf{D2 Index test} & \textbf{D3 Reference method} & \textbf{D4 flow and timing} & \textbf{Overall} \\ \hline

Wang et al., 2016 \cite{b32}	&  Low & Low & 	High & 	Low & 	high \\
Chen et al., 2020 \cite{b2}	& Low & 	Low & 	Low & 	Low & 	low \\
Liu et al., 2020 \cite{b17}	 & Low & 	Low & 	Low & 	Low & 	Low \\
Grzadziel et al., 2020 \cite{b10}	 & High	 & Low	 & High	 & Low	 & High \\
Dai et al., 2021 \cite{b4}	 & Low	 & Low	 & Low	 & Low	 & Low \\
Gao et al., 2021 \cite{b8}	& Low	 & Low	 & Low	 & Low	 & Low \\
Irmak et al., 2021 \cite{b13}	& Low	 & Low	 & Low	 & Low	 & Low \\
Oria et al., 2021 \cite{b21}	& Low	 & Low	 & Low	 & Low	 & Low \\
Rossi et al., 2021 \cite{b24}	& Low	 & Low	 & Low	 & Low	 & Low \\
Xue et al., 2021 \cite{b37}	& Low	 & Low	 & Low	 & Low	 & Low \\
Zhao et al., 2021 \cite{b39}	& Low	 & Low	 & Low	 & Low	 & Low \\
Chen et al., 2022 \cite{b3}	& Low	 & Low	 & Low	 & Low	 & Low \\
Deng et al., 2022 \cite{b5}	& Low	 & Low	 & Low	 & Low	 & Low \\  
O'Hara et al., 2022 \cite{b20}	& Low	 & Low	 & Low	 & Low	 & Low \\
Wang et al., 2022 \cite{b34}	& Low	 & Low	 & Low	 & Low	 & Low \\
Wu et al., 2022 \cite{b35}	& Low	 & Low	 & Low	 & Low	 & Low \\
Allen et al., 2023 \cite{b1}	 & Low	 & Low & Some concerns & Low	 & Low \\
Gao et al., 2023 \cite{b7}	& Low	 & Low	 & Low	 & Low	 & Low \\
Liu et al., 2023 \cite{b16}	& Low	 & Low	 & Low	 & Low	 & Low \\
Suwanraksa et al., 2023 \cite{b27}	& Low	 & Low	 & Low	 & Low	 & Low \\
Szmul et al., 2023 \cite{b28}	& High	& Low	 & Low	 & Low	 & Low	 \\
Wan et al., 2023 \cite{b31}	& Low	 & Low	 & Low	 & Low	 & Low \\
Kong et al., 2023 \cite{b14}	& Low	 & Low	 & Low	 & Low	 & Low \\ 
Wang et al., 2023 \cite{b33}	& Low	 & Low	 & Low	 & Low	 & Low \\
Wynne et al., 2023 \cite{b36}	& Low	 & Low	 & Low	 & Low	 & Low \\
Yoganathan et al., 2023 \cite{b38}	& Low	 & Low	 & Low	 & Low	 & Low \\
Liang et al., 2024 \cite{b15}	& Low	 & Low	 & Low	 & Low	 & Low \\
Peng et al., 2024 \cite{b23}	& Low	 & Low	 & Low	 & Low	 & Low \\ \hline
\end{tabular}
\end{table*}

\newpage

\begin{table*}[!h]
\centering
\begin{tabular}{p{100pt}p{70pt}p{50pt}p{80pt}p{70pt}p{50pt}}
\hline
\textbf{Study} & \textbf{D1 Patient selection} & \textbf{D2 Index test} & \textbf{D3 Reference method} & \textbf{D4 flow and timing} & \textbf{Overall} \\ \hline
Lerner et al. (2021) \cite{b19}	& Low	& Low	& Low	& High	& Low \\ 
Thummerer et al. (2021) \cite{b52}	& Low	& Low	& Low	& High	& Low \\
Pang et al. (2023) \cite {b43}	& Low	& Low	& Low	& Low &	Low  \\
Taasti et al. (2023)\cite {b41}	& Low	& Low	& High	& Low	& Low \\
Thummerer, Zaffino et al. (2020) \cite {b48}	& Low	& Low	& Low	& Low	& Low  \\
Thummerer, de Jong, et al. (2020) \cite {b57}	& High	& Low	& Low	& High	& High  \\
Perez Moreno et al. (2023) \cite {b26}	& High	& Low	& Low	& High	& High \\
\hline
\end{tabular}
\end{table*}

\section{Risk of bias domains conducted using QUADAS 2.}

\begin{figure*}[!h]
\centering
\includegraphics[width=.45\linewidth]{fig1_supp.png}
\label{Quadas2}
\end{figure*}

\newpage

\section*{Acknowledgement}

All authors declare that they have no known conflicts of interest in terms of competing financial interests or personal relationships that could have an influence or are relevant to the work reported in this paper. 


%

\bibliography{refs}